\documentclass[12pt]{article}

\usepackage[dvipdfmx]{color}%{graphicx}
\usepackage{tikz}
\usepackage{amssymb,amsmath}
\usepackage{amsthm}
\usepackage{amsfonts}
\usepackage{comment}
\usepackage{enumerate}
\usepackage{mathrsfs}
\usepackage{ascmac}
\usepackage{color}
\usepackage{cite}
\usepackage[top=30truemm,bottom=30truemm,left=25truemm,right=25truemm]{geometry}
\usepackage[color]{showkeys}
\usepackage{ulem}

\providecommand{\keywords}[1]{\textbf{{\bf Key Words:}} #1}
\providecommand{\Acknowledgement}[1]{\textbf{{\bf Acknowledgement:}} #1}
\let\Re\relax
\DeclareMathOperator{\Re}{Re}
\let\Im\relax
\DeclareMathOperator{\Im}{Im}

\newtheorem{thm}{Theorem}[section]
\newtheorem{prop}[thm]{Proposition}
\newtheorem{lemma}[thm]{Lemma}
\newtheorem{defn}[thm]{Definition}
\newtheorem{rem}[thm]{Remark}

\newtheorem{exam}[thm]{Example}
\newtheorem{cor}[thm]{Corollary}

\title{Spectral mapping theorem of an abstract non-unitary quantum walk}
\author{Keisuke ASAHARA 
	\thanks{The Center for Data Science Education and Research, Shiga University, 1-1-1 Banba hikone, Shiga 522-8522, Japan.
		E-mail: keisuke-asahara@biwako.shiga-u.ac.jp}
	\and Daiju FUNAKAWA 
	\thanks{Department of Electronics and Information Engineering, 
		Hokkai-Gakuen University, Sapporo 062-8605,Japan.
		E-mail: funakawa@hgu.jp}
	\and Etsuo SEGAWA 
	\thanks{Graduate School of Environment and Information Sciences, Yokohama National University, Yokohama 240-8501, Japan.
		E-mail: segawa-etsuo-tb@ynu.ac.jp}
	\and Akito SUZUKI 
	\thanks{Division of Mathematics and Physics,
		Faculty of Engineering,
		Shinshu University,
		Wakasato, Nagano 380-8553, Japan.
		E-mail Address: akito@shinshu-u.ac.jp }
	\and Noriaki TERANISHI
	\thanks{Department of Mathematics, Faculty of Science, Hokkaido University,
		Sapporo 060-0810, Japan.
		E-mail Address: teranishi@math.sci.hokudai.ac.jp
	}
}
\numberwithin{equation}{section}

\begin{document}
	\maketitle
	%\tableofcontents
	%\begin{center}
	%{\}
	%\end{center}
	\begin{abstract}

		This paper continues the previous work
		(Quantum Inf. Process {\bf 11} (2019))
		by two authors 
		of the present paper
		about a spectral mapping property 
		of chiral symmetric unitary operators.  
		In physics, they treat non-unitary time-evolution operators
		to consider quantum walks in open systems. 
		In this paper, we generalize the above result
		to include a chiral symmetric non-unitary operator
		whose coin operator only has two eigenvalues. 
		As a result, the spectra of such non-unitary operators 
		are included in the (possibly non-unit) circle
		and the real axis in the complex plane. 
		We also give some examples of our abstract results,
		such as non-unitary quantum walks defined by Mochizuki {\it et al}.
		Moreover, we present an application to 
		the Ihara zeta functions
		and correlated random walks on regular graphs,
		which are not quantum walks.

	\end{abstract}
	
	\section{Introduction}%section1
	%%%%%%%%%%%%%%%%%%%%%%%%%%%%%%%%%%%%%%%%%%%%%%%%%%%%%%%%%%%%%%%%%%%%%%%%%%%
	%general QW
	%Quantum walks, there are many studies after appearing \cite{Grover.96}, 
	%have interesting researches ones [][][] are applicable to ???? 
	%and \cite{SS.19}[][] are theoretically important. 
	
	%Spectral analysis
	Discrete-time quantum walks (QWs) are quantum counterparts of random walks, and many authors have studied them from several points of view \cite{AsaharaEtAl,RenEtAl.11,Suzuki.16, Suzuki.19, SS.16, SS.19, MKO.16, NOW, KRBD.10, CGGSVWW.18,W19,W20}
	(see \cite{Konno08, V} for reviews). In closed systems, their time-evolution operators are given as unitary operators on the Hilbert spaces of states
	describing their positions and inner degrees of freedom. 
	Then the spectral analysis of QWs gives us rich information such as their long-time behavior \cite{FFS3,Suzuki.16, W20, MSW}, localization\cite{FFS,FFS2,FNSS, Konno2010}, and topological indices\cite{Ma,MST,T, KRBD.10,AO,CGGSVWW.18}. 
	
	On the other hand, several authors have started the study of quantum walks in open systems \cite{MKO.16,RCetal,RMetal,APSS,AGS}. Attal et al. introduced open-quantum random walks to describe such a system using the time-evolution of a density operator\cite{APSS}. Another way to describe such a system is to use non-unitary operators as the time-evolution operators\cite{MKO.16}. Moreover, such operators are not only non-unitary but even non-normal\cite{AsaharaEtAl}. In general, non-normal operators do not satisfy the spectral theorem, making analysis of their spectra difficult. In this paper, we study the spectral analysis of such non-normal time-evolution operators. 
	
	%such as their long-time behavior and localization[].
	%A spectral mapping technique \cite{SS.19} from self-adjoint operators $T$
	%to unitary operators $U$ is know to be useful for analyzing spectra of quantum walks. 
	%\textcolor{red}{Recently, quantum walks in open systems %, 
	%which incorporate the effects of photon inflow and outflow, 
	%have been studied[]. Their time evolution operators are given by non-unitary operators. Moreover this operator is non-normal in general. Therefore, theorems imposing normality, such as the spectral theorem, cannot be used, making analysis difficult. Also, since the time evolution is non-unitary, the probability interpretation is nontrivial. However, open-system quantum walks have been realized and studied through various experiments and simulations in physics[]. Therefore, quantum walks described by non-unitary time evolution operators should be properly formulated in the future.}
	
	Two of authors of this paper showed any chiral symmetric unitary operator has a spectral mapping property from a self-adoint operator by the Joukowsky transform (divided by two) \cite{SS.19, SS.16}. 
	More precisely, a bounded operator $u$ is said to have chiral symmetry
	if there exists a unitary involution $\gamma$,
	{\it i.e.}, $\gamma^{-1} = \gamma^* =\gamma$, %$\gamma^2=1$, 
	such that 
	\[ \gamma u \gamma^\ast 
	= u^\ast. \]
	In particular, a unitary operator $u$ on a Hilbert space $\mathfrak{H}$
	has chiral symmetry 
	if and only if $u$ can be written 
	as a product $u=sc$ of two unitary involutions $s$ and $c$ on $\mathfrak{H}$.
	In this case, there exists a coisometry $d$ from $\mathfrak{H}$
	to a Hilbert space $\mathfrak{K}$ such that
	\[ c = d^*d + (-1) (1 - d^* d), \]
	where $d^*d$ and $1-d^*d$ are the projections onto $\ker (c -1)$
	and  $\ker (c +1)$, respectively \cite{Suzuki.19}. 
	Denoting by $T$ the self-adjoint operator $dsd^*$ on $\mathfrak{K}$,
	we have the spectral mapping property
	\begin{align*}
		& \sigma_{\rm c}(u) \setminus{\{\pm 1\}} 
		= \varphi^{-1}( \sigma_{\rm c} (T)  \setminus{\{\pm 1\}} ), \\
		& \sigma_{\rm p}(u) \setminus{\{\pm 1\}} 
		= \varphi^{-1}( \sigma_{\rm p} (T)  \setminus{\{\pm 1\}} )
		\cup\{+1\}^{m_++M_+} \cup\{-1\}^{m_-+M_-},
	\end{align*} 
	where $\varphi(z) = (z+z^{-1})/2$ is the Joukowsky transform
	and $m_\pm = \dim \ker(T \mp 1)$ and $M_\pm = \dim \ker(s\mp 1)\cap \ker d$. Here we use $\{\lambda \}^m$ to denote that the multiplicity of an eigenvalue $\lambda$ is $m$. 
	For classical results and examples of the spectral mapping property, see \cite{SS.19} and the references. 
	
	This paper reveals the spectral mapping property
	for a chiral symmetric non-unitary bounded operator $U$.
	Let $S$ be a self-adjoint involution and let $d$ be as stated above. 
	Then the chiral symmetric operator $U$ is defined as
	\[ U = S C, \]
	where  
	\[ C = a d^*d + b (1 - d^* d) \quad (a,~ b \in \mathbb{R}),  \]
	is a self-adjoint operator with eigenvalues $a$ and $b$. 
	The chiral symmetry 
	$\gamma U \gamma^* = U^*$ of $U$ can be easily check 
	if we take $\gamma = S$. 
	Our goal of the paper is to state the spectrum mapping property of $U$ by using a scaling Joukowsky transform  $\varphi_{a, b}(z)$ defined in \eqref{sjt}.
	This abstract non-unitary (possibly non-normal) operator $U$
	includes the non-unitary time-evolution operators of quantum walks introduced by Mochizuki, Kim, Obuse \cite{MKO.16}
	and ones of correlated random walks on $k(\geq 2)$-regular graphs.
	Moreover, $U$ is applicable for the positive supports $U^+$, which are non-unitary, of the time-evolution operators of the (unitary) Grover walks on $k(\geq 2)$-regular graphs. In this case, $U^+$ gives a nice expression of the Ihara zeta function \cite{RenEtAl.11}. 
	The organization of the paper is as follows. 
	Sec. 1 briefly introduces the spectral mapping properties and their applications. 
	Sec. 2 gives mathematical preliminaries and main results. 
	Sec. 3 and 4 are the proofs of the spectral mapping properties 
	for eigenvalues and continuous spectra, respectively. 
	In Sec. 5, we give applications of the spectral mapping property. 
	We deal with the Mochizuki-Kim-Obuse model in Sec. 5.1. 
	In Sec. 5.2, we discuss the relation between the positive supports of the Grover walks and the Ihara zeta functions.
	Sec. 5.3 is devoted to the correlated random walks.
	The appendix provides a general property of the resolvent set of $U$ and two examples of our model.
	The two examples presented here are space-homogeneous and space-inhomogeneous quantum walks.

	%%%%%%%%%%%%%%%%%%%%%%%%%%%%%%%%%%%%%%%%
	%%%%%%%%%%%%%%%%%%%%%%%%%%%%%%%%%%%%%%%%
	\section{Preliminaries and Main Theorem}
	%%%%%%%%%%%%%%%%%%%%%%%%%%%%%%%%%%%%%%%%
	%%%%%%%%%%%%%%%%%%%%%%%%%%%%%%%%%%%%%%%%
	%In this section, we will first prepare some notations and then introduce the main result.
	
	%%%%%%%%%%%%%%%%%%%%%%%%%%
	\subsection{Preliminaries}
	%%%%%%%%%%%%%%%%%%%%%%%%%%
	Let $\mathcal{H}$ and $\mathcal{K}$ be (non-trivial) complex Hilbert spaces. We use $d$ to denote a coisometry from $\mathcal{H}$ to $\mathcal{K}$, i.e., $d$ is a bounded linear operator and satisfies 
	%Let $\mathcal{H}$ and $\mathcal{K}$ be complex Hilbert spaces with $\dim\mathcal{K}\geq 1$ and 
	%$d$ be a coisometry from $\mathcal{H}$ to $\mathcal{K}$, in other words, it is a bounded linear operator and satisfies 
	\begin{align}
		\label{eq:dada*=I}
		d d^\ast=I_{\mathcal{K}}, 
	\end{align}
	where $I_{\mathcal{N}}$ is the identity operator on a Hilbert space $\mathcal{N}$.
	The operator $d$ is the same one as $d_{\mathcal{A}}$ in \cite{SS.19}.
	For given $a,b\in \mathbb{R}$, we define a bounded self-adjoint operator $C$ on $\mathcal{H}$ as follows
	\begin{align*}
		C:=& \ a d^\ast d + b (I_{\mathcal{H}} - d^\ast d)\\
		=& \ (a-b)d^\ast d + b I_{\mathcal{H}}
	\end{align*}
	and call it a coin operator.
	
	Let us recall the basic properties of $d$ and $C$. 
	See \cite{SS.19} for detail.
	%Here, we confirm $d$ and $C$.
	We will use the symbol $\sigma(A)$ and $\sigma_\mathrm{p}(A)$ to denotes the spectrum and point spectrum of a 
	linear operator $A$ respectively. 
	We observe that $d^\ast d$ is a non-zero orthogonal projection on $\mathcal{H}$ because $d^\ast$ is injective and $d$ is surjective. 
	Moreover, the following lemma holds. 
	
	%%% Property of d^\ast d and C %%%
	\begin{lemma}
		The spectrum of $C$ consists only of its eigenvalues and that is contained by $\{a,b\}$.
		Moreover, the following form holds
		\begin{align}
			\label{eq:sigmaC}
			\sigma(C)=\sigma_\mathrm{p}(C)=\{a,b\}
		\end{align}
		if and only if $d^\ast d\neq I_{\mathcal{H}}$.
	\end{lemma}
	\begin{proof}
		If $d^\ast d = I_{\mathcal{H}}$, then $C=a I_{\mathcal{H}}$ and so 
		$\sigma(C) = \sigma_\mathrm{p}(C) = \{a\}$. 
		If $d^\ast d \neq I_{\mathcal{H}}$, then \eqref{eq:sigmaC} holds since $d^\ast d$ and $I_{\mathcal{H}} - d^\ast d$ are non-zero and orthogonal to each other.
	\end{proof}

	\begin{comment}
	\begin{proof}
	(i)
	\cite[Preliminaries(2.1)]{SS.19} no sita $\rightarrow$ $\Pi_{\mathcal{A}}$ is a projection.\\
	We prove that it is not zero operator.
	By the definition of $d$, it is surjective and $d^\ast$ is injective. 
	There exists $\psi \in \mathcal{H}$ such that 
	$d\psi \neq 0_{\mathcal{K}}$. 
	It holds that $\Pi_{\mathcal{A}}\psi \neq 0_{\mathcal{H}}$ since $d^\ast$ is injection. Hence $\Pi_{\mathcal{A}}$ is not zero.\\
	(ii)
	By the definition of $C$ and $\Pi_{\mathcal{A}}$ be a projection, it follows that
	\[C=aI\oplus bI\]
	on an orthogonal decomposition 
	$\mathcal{H}={\rm Ran}(\Pi_{\mathcal{A}})\oplus {\rm Ran}(I_{\mathcal{H}} - \Pi_{\mathcal{A}})$.
	In general, spectrum and point spectrum of direct sum of linear operators are given by the union of each operator's spectrum, so we get the equation \eqref{eq:sigmaC}.
	\end{proof}
	\end{comment}
	
	Let $S$ be a unitary involution, which is hence a self-adjoint operator, on $\mathcal{H}$ and we call it a shift operator. 
	We observe that $S^2=I_{\mathcal{H}}$ holds from $S^\ast = S$. 
	Moreover, $dS$ is a coisometry from $\mathcal{H}$ to $\mathcal{K}$ because so is $d$ and $S^2=I_{\mathcal{H}}$.
	Our model includes that in \cite{SS.19} in the case $a=1, b=-1$.

	%%% Property of d_B^\astd_B and S %%%
	\begin{rem}
		The spectrum of $S$ consists only of its eigenvalues and that is contained by $\{\pm1\}$.
		Moreover, $\sigma(S) = \sigma_\mathrm{p}(S) =\{\pm1\}$ if and only if 
		$S\neq \pm I_{\mathcal{H}}$.
	\end{rem}
	\begin{comment}
	\begin{proof}
	The spectrum of $S$ is included by unit circle and real number because $S$ is unitary and self-adjoint, respectively. 
	Thus we have $\sigma(S) \subset \{\pm 1\}$. 
	$\sigma_\mathrm{p}(S) \subset \sigma(S)$
	\end{proof}
	\end{comment}    
	
	%definition of U and T
	\begin{defn}\label{def.2.3}
		We define a time evolution operator $U$ associated with $S$ and $C$ by 
		\begin{equation*}
			U:=SC.
		\end{equation*} 
		The discriminant operator $T$ of $U$ is also defined by
		\begin{equation*}
			T:= dS d^\ast.
		\end{equation*} 
	\end{defn}
	This time-evolution operator is simple but covers a lot of previous unitary and non-unitary walk models. For examples, this time evolution operator can reproduce not only a standard time evolution operator of a discrete-time quantum walk, for example, the split step model~\cite{KRBD.10} and the Szegedy walk on a graph~\cite{EmmsEtAl.06,Szegedy.04}, which are unitary, but also the Mochizuki-Kim-Obuse model~\cite{MKO.16} and the Bass-Hashimoto expression of the Ihara zeta function~\cite{Bass.92,Hashimoto.89}, which are non-unitary (see also examples in Appendix). Moreover this operator $U$ also reproduces a correlated random walk on a graph~\cite{RH,Konno.09}.
	
	The operator $U$ is a bounded operator on $\mathcal{H}$ since so is $S$ and $C$. 
	We note that $T$ is a self-adjoint operator equipped with its operator norm $\|T\|\leq 1$ (see \cite{SS.19} for details).
	Moreover, it has chiral symmetry from $SUS=U^\ast$.
	\begin{rem}
		From the definition of $U$, 
		\begin{equation}
			\label{uu*}
			[U, U^\ast ] = \left(a^2 -b^2\right)[S, d^\ast d]S
		\end{equation}
		holds, where $[A, B]:=AB-BA$ denotes the commutator of linear operators $A$ and $B$. 
		Because of the unitarity of $S$, we see that $U$ is normal if and only if 
		$a^2 = b^2$ or $[S, d^\ast d ]= 0$. 
		The equation \eqref{uu*} implies that U is not always normal.
		Note that spectral theory is not applicable to non-normal case, especially to non-unitary case. 
		See Example %$\ref{example1}$, 
		$\ref{example2}$ and $\ref{example3}$ for non-unitary cases. 
	\end{rem}
	
	A rough estimate of the range of $\sigma(U)$ is immediately obtained from the definition of $U$.
	\begin{cor} \label{cor.2.5}
		The spectrum of $U$ is included in 
		\begin{equation*}
			\bigl\{z\in \mathbb{C} \ \big|\ \min\{|a|,|b|\} \leq |z| \leq \max\{|a|,|b|\} \bigr\}.
		\end{equation*}
	\end{cor}
	
	This follows from Proposition \ref{rho(U)}, which is proven in the appendix. 
	In particular, we emphasize that $\sigma(U)$ does not include the origin if $ab \neq 0$ and the real part of $\sigma(U)$ falls in the range 
	\begin{equation*}
		[-\max\{|a|,|b|\}, \ -\min\{|a|,|b|\}] \cup
		[\min\{|a|,|b|\}, \ \max\{|a|,|b|\}].
	\end{equation*}

	%%%%%%%%%%%%%%%%%%%%%%%%%
	\subsection{Main Theorem}
	%%%%%%%%%%%%%%%%%%%%%%%%%
	%In this subsection, we introduce our conclusion. 
	
	%%% Assumption %%%
	In what follows, we formulate our assumptions.
	If $d^\ast d= I_{\mathcal{H}}$ (respectively, $a=b$, $S=\pm I_{\mathcal{H}}$), 
	then $U=a S$ (resp., $U=bS$, $U=\pm C$), so we get the spectrum of $U$ from that of $S$ (resp., $S$, $C$) immediately.
	Moreover, if $a=-b$, then $C=a(2d^\ast d-I_{\mathcal{H}})$, so the spectrum of $U$ is provided by \cite{SS.19} immediately. 
	
	%spectral property
	\begin{prop}
		Assume $d^*d \not= I_\mathcal{H}$, $S \not=\pm I_\mathcal{H}$,and $a\not=\pm b$.  
		The residual spectrum of $U$ is empty. 
	\end{prop}
	This is proved by Remark \ref{sigmapsymm} and the chiral symmetry of $U$. 
	Hence, the spectrum of $U$ consists only of the point spectrum and continuous spectrum:
	\[\sigma(U) = \sigma_\mathrm{p}(U)\cup \sigma_\mathrm{c}(U),\]
	where $\sigma_\mathrm{c}(A)$ denotes the continuous spectrum of a linear operator $A$. 
	Generally speaking, the residual spectrum of a linear operator is empty when it is unitary, or more generally, normal. 
	Because $U$ is not unitary in our paper,
	the absence of residual spectrum plays a key role
	in the proof of the continuous spectrum part of our main theorem stated later.
	
	Our goal of the present paper is to state the spectrum mapping property of $U$ by using a scaling Joukowsky transform $\varphi_{a,b}$ defined as follows:  
	Recall that the Joukowsky transform is given by 
	$\varphi(z):=(z+z^{-1})/2$ for $z\in\mathbb{C}\setminus \{0\}$. Then we define the scaled Joukowsky transform 
	$\varphi_{a,b} {\color{red}\colon} \mathbb{C}\setminus \{0\}\rightarrow \mathbb{C}$ as 
	\begin{equation}
		\label{sjt}
		\varphi_{a,b}(z):=\frac{2\sqrt{-ab}}{a-b}\varphi\left(\frac{z}{\sqrt{-ab}}\right)
		=\frac{z-ab z^{-1}}{a-b}, \quad z\in \mathbb{C}\setminus\{0\}.
	\end{equation}

	Further, if $ab=0$, $U$ is not invertible, and $0$ is included in the spectrum of $U$. In this case, the spectrum of $U$ is not included in the domain of the scaled Joukowsky transform. Hence it is natural to assume that $ab\not=0$.
	Therefore, we prove our main theorem under the following conditions:
	\begin{align}\label{ass1}
		d^\ast d\neq I_{\mathcal{H}}, \quad S\neq \pm I_{\mathcal{H}}, 
		\quad a\neq \pm b, \quad ab\neq0.
	\end{align}
	
	Throughout this paper, except for the appendix, 
	we assume that the above conditions \eqref{ass1} hold.
	We are now in a position to state our main result.

	%To mention our main conclusion, we introduce a scaled Joukowsky transform 
	%$\varphi_{a,b} {\color{red}\colon} \mathbb{C}\setminus \{0\}\rightarrow \mathbb{C}$ by using the Joukowsky transform $\varphi(z):=(z+z^{-1})/2$ {\color{red}for }$z\in\mathbb{C}\setminus \{0\}$ as follows:
	
	\begin{thm}\label{Main}
		The followings hold:
		\begin{align}
			&\sigma_\mathrm{c}(U)\setminus \{\pm a, \pm b\} = \varphi_{a,b}^{-1}(\sigma_\mathrm{c}(T)\setminus\{\pm1\} ),\\
			&\sigma_\mathrm{p}(U)=\varphi_{a,b}^{-1}\bigl(\sigma_\mathrm{p}(T)\setminus \{\pm1\}\bigr) 
			\cup\{a\}^{m_{+} }\cup\{-a\}^{m_{-} }\cup\{-b\}^{M_{+} }\cup\{b\}^{M_{-} }.\label{Main:eqpoint}
		\end{align}
		If $\pm a\in \sigma_\mathrm{c}(U)$, then $\pm1\in\sigma_\mathrm{c}(T)$ and if $\mp b \in \sigma_\mathrm{c}(U)$, then $\pm1\in\sigma(T)$ respectively.
		If $\pm 1 \in \sigma_\mathrm{c}(T)$, then 
		$\pm a \in \sigma_\mathrm{c}(U),\ \mp b\in\sigma(U)$ respectively.
		Moreover, for any $\lambda\in\sigma_\mathrm{p}(U)$, 
		\begin{align}
			\label{eq:Main,sigmapU}
			\dim \ker (U-\lambda)=\left\{
			\begin{array}{ll}
				\dim \ker (T-\varphi_{a,b}(\lambda)), & \lambda\neq \pm a,\ \pm b,\\
				m_{+}, & \lambda=a,\\
				m_{-}, & \lambda=-a,\\
				M_{+}, & \lambda=-b,\\
				M_{-}, & \lambda=b,
			\end{array}
			\right.
		\end{align}
		where 
		\begin{align}
			m_{\pm}:=\dim \ker(T\mp1),\quad 
			M_{\pm}:=\dim \left[\ker d  \cap \ker(S\pm1)\right].
			\label{Main:dim}
		\end{align}
	\end{thm}
	
	\begin{center}
		%\documentclass{article}
%\usepackage[dvipdfmx]{color}%{graphicx}
%\usepackage{tikz}
%\begin{document}
\begin{tikzpicture}

    %real axis
    \draw [dotted,-stealth](-3.5,0)--(3.6,0);
    %circle dotted
    \draw [dotted, domain=-2:2, samples=200] plot(\x, {(4-(\x)^2)^(0.5)});
    \draw [dotted, domain=-2:2, samples=200] plot(\x, {-(4-(\x)^2)^(0.5)});
    %circle red
    \draw [very thick, red, domain=-1:2, samples=200] plot(\x, {(4-(\x)^2)^(0.5)});
    \draw [very thick, red, domain=-1:2, samples=200] plot(\x, {-(4-(\x)^2)^(0.5)});
    %circle supplement
    \draw [very thick, red] (2,0.15)--(2,0);
    \draw [very thick, red] (2,-0.15)--(2,0);
    %arrow
    \draw [dotted,-stealth](0.2,0.1)--(1.45,1.25);
    \draw [dotted,-stealth](0.2,-0.1)--(1.45,-1.25);
    \draw [dotted,-stealth](-0.5,0.1)--(-0.95,1.68);
    \draw [dotted,-stealth](-0.5,-0.1)--(-0.95,-1.68);
   %node a
   \fill [red](-1.2,0) circle (2pt);
   \node [anchor=north] at (-1.3,-0.1){$-a$};
   \fill [red](1.2,0) circle (2pt);
   \node [anchor=north] at (1.3,-0.1){$a$};
   %node b
   \fill [red](-3,0) circle (2pt);
   \node [anchor=north] at (-2.9,-0.1){$b$};
   \fill [red](3,0) circle (2pt);
   \node [anchor=north] at (2.8,-0.1){$-b$};
    %real spectrum of U
    \draw [very thick, red](1.4,0)--(2.8,0);
    %spectrum of T
    \draw [very thick](-0.5,0)--(0.8,0);
    
    %words
    \node [anchor=south] at (0.4, 0.6) {$\varphi_{a,b}^{-1}$};
    \node [anchor=north] at (0.4, -0.6) {$\varphi_{a,b}^{-1}$};
    \node [anchor=north] at (0,-0.1) {$\sigma(T)$};
    \node [anchor=south, red] at (2,1.5){$\sigma(U)$};
    \node [anchor=east] at (0,-2.5){Radius: $\sqrt{-ab},$};
    \node [anchor=west] at (0.2,-2.5){Birth: $\pm b$};
    \node [anchor=north] at (0,-2.7){In the case $1<a<-b$};
    \end{tikzpicture}
    %\caption{Figure}
%\end{document}

		%\caption{Figure}
		\label{Figure}
	\end{center}

	\begin{rem}
		Theorem $\ref{Main}$ shows that the point and continuous spectrum of $U$ except for $\{\pm b\}$ corresponds respectively to that of $T$ via $\varphi_{a,b}$. This is one of the differences between our results and {\rm \cite{SS.19}}. 
		Also, in the previous study, there are birth eigenvalues $\pm 1$ as an eigenvalue of $U$ that does not originate from $T$. The birth eigenvalue in the present study is $\pm b$. Thus, an eigenvalue of the coin operator for $1-d^*d$ correspond to the  birth eigenvalues.
		
		For $\lambda\in\sigma(U)$, let $t=\varphi_{a,b}(\lambda)$ then the following equation holds:
		\begin{align}\label{eq:lambda=ft}
			\lambda = \frac{(a-b)t\pm\sqrt{(a-b)^2 t^2 + 4ab} }{2}.
		\end{align}
		Recall that $t$ is a real number since $T$ is self-adjoint. 
		This equation indicates that $\sigma(U)$ becomes a subset of real axis if $ab>0$.
		Moreover, it may have its imaginary part if $ab<0$, and
		$|\lambda|=\sqrt{-ab}$ holds by \eqref{eq:lambda=ft}
		for $\lambda$ satisfying $\Im \lambda \neq 0$. 
		Hence we obtain $\sigma(U)\subset (\mathbb{R}\cup \sqrt{-ab}\mathbb{T})$.
	\end{rem}

	%%% notations %%%
	To show that Theorem \ref{Main} holds, we provide several notations:
	\begin{align}
		&\mathcal{L}:={\rm ran}(d^\ast d)+{\rm ran}(Sd^\ast dS),\nonumber \\
		&\mathcal{L}_1:=
		d^\ast\bigl(\ker(T^2-1)^{\perp}\bigr) + (dS)^\ast \bigl(\ker(T^2-1)^{\perp}\bigr),\nonumber\\
		\label{eq:L0def}
		&\mathcal{L}_0^{\pm}:=d^\ast \ker(T\mp 1),\\
		&\mathcal{L}_0:=\mathcal{L}_0^{+} \oplus \mathcal{L}_0^{-},\nonumber\\
		&\mathcal{L}_{\pm}^{\perp}:=\mathcal{L}^{\perp} \cap \ker(S\pm1),\nonumber
	\end{align}
	where $E+F=\{e+f \mid e \in E, \ f \in F \}$,
	%denotes the all vector represented by sum of elements in $E$ and $F$ of subsets $E$ and $F$ contained in a vector space, 
	and $M^\perp$ denotes the orthogonal complement of a subspace $M$.
	
	%%%%%%

	%%%%%%%%%%%%%%%%%%%%%%%%%%%%%%%%%%%%%%%%%%%%%%%%%%%%%%%%%%%%%%%%%%%%%%%%%
	\section{Eigenvalues of time evolution}
	%%%%%%%%%%%%%%%%%%%%%%%%%%%%%%%%%%%%%%%%%%%%%%%%%%%%%%%%%%%%%%%%%%%%%%%%%
	In this section, we will prove the next theorem.
	\begin{thm}\label{Maineigenvalue}
		The point spectrum of $U$ is given by the following:
		\begin{align}
			\label{eq:sigmapU}
			\sigma_\mathrm{p}(U)=\varphi_{a,b}^{-1}(\sigma_\mathrm{p}(T)\setminus \{\pm1\}) 
			\cup\{a\}^{m_{+} }\cup\{-a\}^{m_{-} }\cup\{-b\}^{M_{+} }\cup\{b\}^{M_{-} }.
		\end{align}
		Moreover, for any $\lambda\in\sigma_\mathrm{p}(U)$, 
		\begin{align*}
			\dim \ker (U-\lambda)=\left\{
			\begin{array}{ll}
				\dim \ker (T-\varphi_{a,b}(\lambda)), & \lambda\neq \pm a, \pm b,\\
				m_{+}, & \lambda=a,\\
				m_{-} , & \lambda=-a,\\
				M_{+}, & \lambda=-b,\\
				M_{-}, & \lambda=b.
			\end{array}
			\right.
		\end{align*}
	\end{thm}

	\begin{rem}\label{sigmapsymm}
		Theorem $\ref{Maineigenvalue}$ indicates that 
		$\sigma_\mathrm{p}(U)$ is reflection symmetry with respect to the real axis.
		Actually, for all $\lambda\in \sigma_\mathrm{p}(U),\varphi_{a,b}(\lambda)$ is a real number and hence 
		\begin{align*}
			\frac{\lambda-ab \lambda^{-1}}{a-b}
			=\frac{\overline{\lambda}-ab \overline{\lambda}^{-1} }{a-b}
		\end{align*}
		holds, where $\overline{\eta}$ is the complex conjugate of $\eta\in\mathbb{C}$. This implies $\overline{\lambda}\in\sigma_\mathrm{p}(U)$ by \eqref{eq:sigmapU}. 
	\end{rem}
	
	\begin{comment}
	\begin{rem}
	From $\sigma_\mathrm{p}(U) = \sigma_\mathrm{p}(U)^\ast$ 
	\eqref{sigmaW=sigmaW*}, we have that if 
	$\lambda$ is an eigenvalue of $U$, so is $\overline{\lambda}$, 
	where $\overline{c}$ denotes the complex conjugate of $c\in\mathbb{C}$.
	\end{rem}
	\end{comment}

	\begin{comment}
	$d^\ast f^{\pm}\in \mathcal{L}_0^{\pm}$, 
	\begin{align}
	S(d^\ast f^{\pm})=\pm d^\ast f^{\pm}, 
	\quad f^{\pm}\in \ker(T\mp 1),
	\end{align}
	holds, i.e.,
	\end{comment}
	\begin{comment}
	It is easy to see that $\mathcal{L}$ and $\mathcal{L}_1$ are subspaces
	of $\mathcal{H}$.
	Moreover, 
	$\mathcal{L}_0$ and $\mathcal{L}_0^{\pm}$ are closed subspaces of $\mathcal{H}$
	since  $\ker(T^2-1)$ and $\ker(T\pm1)$ are closed subspaces and $d^\ast$ is bounded.   
	The orthogonal decomposition 
	$\mathcal{L}_0=\mathcal{L}_0^{+} \oplus \mathcal{L}_0^{-}$ 
	holds by the orthogonal decomposition $\ker(T^2-1)=\ker(T-1)\oplus \ker(T+1)$ and \eqref{eq:dada*=I}. 
	\end{comment}
	\begin{comment}
	\begin{align*}
	\psi =\psi_0 + d^\ast f.
	\end{align*}
	\end{comment}
	
	The following proposition is important in the sense that it determines the distribution of the spectrum of $U$.
	
	\begin{prop}\label{Tf=phif}
		For $\lambda \in \sigma_\mathrm{p}(U)$ and 
		$\psi \in \ker(U - \lambda)\setminus\{0\}$, then
		\begin{align}\label{eq:Tf=phif}
			Tf=\varphi_{a,b}(\lambda)f,\quad f=d\psi.
		\end{align}
		Moreover if $\lambda\neq\pm b$, then $\varphi_{a,b}(\lambda)\in\sigma_\mathrm{p}(T)$ and it holds that either
		\begin{align}
			\label{eq:condition_lambda}
			\Im\lambda=0 \text{ or } |\lambda|=\sqrt{-ab}.
		\end{align}
	\end{prop}
	
	\begin{comment}
	\begin{proof}
	Since $Cd^\ast f=ad^\ast f$ and $C\psi_0=b\psi_0$, we see that
	\begin{align*}
	\lambda(\psi_0+d^\ast f)&=U\psi\\
	&=SC(\psi_0+d^\ast f)\\
	&=S(b\psi_0+ad^\ast f)\\
	&=bS\psi_0+aSd^\ast f.
	\end{align*}
	Thus, we have 
	\begin{align}
	\label{eq:f-psi0}
	(aSd^\ast-\lambda d^\ast)f=(\lambda-bS)\psi_0.
	\end{align}
	Letting $d$ and $dS$ act on \eqref{eq:f-psi0}, we have equations:
	\begin{align}
	\label{eq:Tf-psi0}
	(aT-\lambda)f&=-b dS\psi_0,\\
	\label{eq:Tf-rpsi0}
	(a-\lambda T)f&=\lambda dS\psi_0. 
	\end{align}
	From the above equations 
	\begin{align*}
	(aT-\lambda)f&=-bdS\psi_0\\
	&=-b(a\lambda^{-1}-T)f.
	\end{align*}
	Hence, we have 
	\begin{equation}
	\label{eq:Tf}
	Tf=\frac{\lambda-ab\lambda^{-1}}{a-b}f=\varphi_{a,b}(\lambda)f.
	\end{equation}
	\end{proof}
	\end{comment}

	\begin{proof}
		We have \eqref{eq:Tf=phif} in the same way as 
		\cite[Proposition 5.1(1)]{SS.19}. 
		Let $\lambda\neq\pm b$. If $f=0$, then we obtain $(\lambda-bS)\psi = 0$ from 
		$U\psi = \lambda \psi$. 
		But this contradicts to $\psi\neq0$ or $\lambda\neq\pm b$. 
		Therefore $f\neq0$ and $\varphi_{a,b}(\lambda)\in\sigma_\mathrm{p}(T)$. 
		Moreover, $\varphi_{a,b}(\lambda)$ becomes a real number since 
		$T$ is self-adjoint, which provides 
		\begin{equation*}
			\left(\lambda - \overline{\lambda}\right)\bigl(|\lambda|^2 - ab \bigr)=0,
		\end{equation*}
		hence we have \eqref{eq:condition_lambda}.
	\end{proof}

	%%% Fact %%%
	\begin{rem}
		Here we introduce some facts without proof because they can be proved by the similar way in {\rm \cite{SS.19}}.
		\begin{align}
			%\label{eq:Hdecomp}
			%\mathcal{H} = \overline{\mathcal{L}} \oplus \mathcal{L}^{\perp},\\
			&\mathcal{L}^{\perp}=\ker d \cap \ker (dS).\\
			\label{eq:Ldecomp}
			&\mathcal{L} = 
			\mathcal{L}_1 \oplus \mathcal{L}_0^{+} \oplus \mathcal{L}_0^{-},\\
			\label{L0subS}
			&\mathcal{L}_0^{\pm} \subset \ker(S\mp1),\\
			\label{eq:directdecompositionH}
			&\mathcal{H}=
			\mathcal{L}^{\perp}\oplus\overline{\mathcal{L}_1}
			\oplus\mathcal{L}_0,
		\end{align}
		where $\overline{M}$ denotes the closure of a set $M$. 
		Moreover, $U$ is reduced by each closed subspace $\mathcal{L}^{\perp}, \overline{\mathcal{L}_1}$ and $\mathcal{L}_0$. 
		Hence the spectrum of $U$ is obtained from the each that of the reduced part of $U$.
		For a closed subspace $M$ in a Hilbert space $\mathcal{N}$ and a linear operator $A$ on $\mathcal{N}$, 
		let $A_{M}$ denote the restriction of $A$ to $M$ if $A$ is reduced by $M$. 
		
		The restriction of $d^\ast$ on $\ker(T\mp1)$ is bijective to $\mathcal{L}_0^\pm$ and satisfies
		\begin{align}
			\label{eq:surj_d}
			\left(d^\ast |_{\ker(T\mp1)}\right)^{-1}
			=d |_{\mathcal{L}_{0}^{\pm}}.
		\end{align}
		
		%\begin{comment}
		%\begin{proof}
		%Bijectivity of $d^\ast|_{\ker(T\mp1)}$ is trivial. For any $g\in\mathcal{L}_0^\pm$, it is clear that $dg\in\ker(T\mp1)$ and 
		%\begin{equation*}
		%	d^\ast dg = g.
		%\end{equation*}
		%The above equation gives us the relation \eqref{eq:surj_d}.
		%\end{proof}
		%\end{comment}
		
		The operators $U_{\mathcal L_0}$ and $U_{\mathcal{L}^\perp}$ are decomposed as follows:
		\begin{align}
			\label{eq:decomposeL0andLperp}
			U_{\mathcal{L}_0}=\left(a I_{\mathcal{L}_0^{+}}\right) 
			\oplus \left(-a I_{\mathcal{L}_0^{-}}\right),\quad 
			U_{\mathcal{L}^{\perp }}
			=\left(-b I_{\mathcal{L}_{+}^{\perp} }\right) 
			\oplus \left(b I_{\mathcal{L}_{-}^{\perp} }\right).
		\end{align}
	\end{rem}    

	\begin{prop}\label{pointspectrum}
		The following relations hold:
		\begin{enumerate}[{\rm (i)}]
			\item 
			%$\sigma_\mathrm{p}(U)\setminus \{\pm a, \pm b\}\subset\sigma_\mathrm{p}(U_{\overline{\mathcal{L}_1 }})\cap \left\{\lambda \in \mathbb{C}\ \middle| \ \varphi_{a,b}(\lambda)\in \sigma_\mathrm{p}(T)\setminus\{\pm1\}\right\}$.
			$\sigma_\mathrm{p}(U)\setminus \{\pm a, \pm b\}\subset\sigma_\mathrm{p}(U_{\overline{\mathcal{L}_1 }})\cap \varphi_{a,b}^{-1}(\sigma_\mathrm{p}(T)\setminus\{\pm1\})$.
			\begin{comment}
			\item 
			$\sigma_\mathrm{p}(U)\setminus \mathbb{R} \subset \sigma_\mathrm{p}(U_{\overline{\mathcal{L}_1 }})\cap \left\{\lambda \in \mathbb{C}\setminus \mathbb{R}\ \middle| \ \dfrac{2}{a-b}\Re \lambda\in \sigma_\mathrm{p}(T)\mbox{ and } |\lambda|=\sqrt{-ab}\right\}$.
			\end{comment}
			\item 
			$\sigma_\mathrm{p}(U) \cap \{\pm a\} \subset \sigma_\mathrm{p}(U_{\mathcal{L}_0 })\cup \sigma_\mathrm{p}(U_{\mathcal{L}^{\perp}})$.
			\item 
			$\sigma_\mathrm{p}(U) \cap \{\pm b\} \subset\sigma_\mathrm{p}(U_{\mathcal{L}_0 })\cup \sigma_\mathrm{p}(U_{\mathcal{L}^{\perp}})$.
			\begin{comment}
			\item 
			If $ab<0$, then
			\begin{align*}
			\sigma_\mathrm{p}(U) \cap \{\pm \sqrt{-ab}\} \subset 
			\sigma_\mathrm{p}(U_{\overline{\mathcal{L}_1 }})\cap 
			\left\{\lambda \in \mathbb{R}\ \middle| \ 
			\frac{2\lambda}{a-b}\in \sigma_\mathrm{p}(T), |\lambda|=\sqrt{-ab}\right\}.
			\end{align*}
			\item 
			\begin{align*}
			\label{U_p R}
			(\sigma_\mathrm{p}(U)\cap\mathbb{R})\setminus \{\pm a, &\pm b, \pm \sqrt{-ab}\}\\
			&\subset 
			\sigma_\mathrm{p}(U_{\overline{\mathcal{L}_1 }})\cap 
			\left\{\lambda \in \mathbb{R}\setminus\{0\} \ \middle| \ 
			\varphi_{a.b}(\lambda)\in \sigma_\mathrm{p}(T), |\lambda|\neq |a|,|b|\right\}.
			\end{align*}
			\end{comment}
		\end{enumerate}
	\end{prop}

	\begin{proof}
		The assertion (i)-(iii) is proved by 
		Proposition \ref{Tf=phif} and a 
		similar way in \cite[Proposition 5.1]{SS.19} which is based on to give the following relations, 
		for any $\lambda\in\sigma_\mathrm{p}(U)$,
		\begin{align}\label{eq:kernelU_inclusion}
			\ker(U-\lambda) \subset \left\{
			\begin{array}{ll}
				\mathcal{L}_1, & \mathrm{if} \ \lambda\neq \pm a, \pm b, \\
				\mathcal{L}_0 \oplus \mathcal{L}^\perp, & \mathrm{if} \ \lambda =  \pm a, \pm b.
			\end{array}
			\right.
		\end{align}
		
	\end{proof}

	\begin{rem}\label{abNotinSigmaUL1}
		The relation \eqref{eq:kernelU_inclusion} implies 
		$\pm a, \pm b \notin \sigma_\mathrm{p}(U_{\overline{\mathcal{L}_1 }})$. 
		Thus Proposition $\ref{pointspectrum}$ gives
		\begin{align*}
			\sigma_\mathrm{p}(U)\setminus \{\pm a, \pm b\}=\sigma_\mathrm{p}(U_{\overline{\mathcal{L}_1 }}),
		\end{align*}
		in particular, $\pm a, \pm b\notin \sigma_\mathrm{p}(U_{\overline{\mathcal{L}_1 }})$.
	\end{rem}

	%%%%%%%%%%%%%%%%%%%%%%%%%%%%%%%%%%%%%%%%%%%%%%%%%%%%%%%%%%%%%%%%%%%%%%%%%%%
	\begin{prop}\label{pointUL1}
		The equation
		\begin{align}
			\label{eq:sigmapUL1=phiinverse}
			\sigma_\mathrm{p}(U_{\overline{\mathcal{L}_1} })
			=\varphi_{a,b}^{-1}(\sigma_\mathrm{p}(T) \setminus \{\pm1\})
		\end{align}
		holds and for any $\lambda\in\sigma_\mathrm{p}(U_{\overline{\mathcal{L}_1} })$, we obtain 
		\begin{align}
			\label{eq:sigmapUL1=phiinverse:dimension}
			\dim \ker(U-\lambda)= \dim \ker(T-\varphi_{a,b}(\lambda)).
		\end{align}
	\end{prop}
	
	%{\it Proof of \eqref{eq:sigmapUL1=phiinverse}. }\\
	\begin{proof}[Proof of the equation \eqref{eq:sigmapUL1=phiinverse}]
		By Proposition \ref{pointspectrum}, the left hand side of \eqref{eq:sigmapUL1=phiinverse} is contained in that of right hand side. 
		Hence it is sufficient to show that the inverse inclusion relation holds.
		For each $\lambda\in\mathbb{C}\setminus\{0\}$, a map $K_{\lambda}$ from $\mathcal{K}$ to $\mathcal{H}$  which is defined by 
		\begin{align*}
			K_{\lambda}:=(b+\lambda S)d^\ast
		\end{align*}
		is a bounded operator. 
		If $\lambda\neq \pm b$, then $b+\lambda S$ is bijection because 
		$b\lambda^{-1}$ is in the resolvent set of $S$.
		In this case, 
		\begin{align}\label{eq:Klambda_invertible}
			d(b+\lambda S)^{-1}K_{\lambda}=I_{\mathcal{K}}
		\end{align}
		holds. 
		Let $\lambda\in \varphi_{a,b}^{-1}(\sigma_\mathrm{p}(T) \setminus \{\pm1\})$.
		Then $\varphi_{a,b}(\lambda)\in \sigma_\mathrm{p}(T)\setminus \{\pm1\}$ holds and 
		for an eigenvector $g\in \ker(T-\varphi_{a,b}(\lambda))\setminus \{0\}$ we define 
		$\phi:=K_{\lambda}g$.  
		The vector $g$ is in $\ker(T^2-1)^{\perp}$ since $\varphi_{a,b}(\lambda) \neq \pm1$ and the eigenspaces of a self-adjoint operator associated with different eigenvalues are orthogonal each other, thus $\phi\in \mathcal{L}_1$. 
		By the definition of $\varphi_{a,b}$, we have $\varphi_{a,b}(\theta)=1$ (resp.\ $-1$) if and only if $\theta=a, -b$ (resp.\ $-a$, $b$), since $\lambda \neq \pm b$. 
		Therefore, we can show that $\phi$ is a non-zero vector since $K_{\lambda}$ is injective from \eqref{eq:Klambda_invertible}. 
		Moreover, the equation $U\phi=\lambda \phi$ holds by $Tg=\varphi_{a,b}(\lambda)g$. 
		This provides us 
		$\sigma_\mathrm{p}(U_{\overline{\mathcal{L}_1} })
		\supset \varphi_{a,b}^{-1}(\sigma_\mathrm{p}(T) \setminus \{\pm1\})$, hence
		\eqref{eq:sigmapUL1=phiinverse} holds. %$\hfill \Box$\\
	\end{proof}
	To get \eqref{eq:sigmapUL1=phiinverse:dimension}, we show that the following lemma holds.

	\begin{lemma}
		For $\lambda\in\sigma_\mathrm{p}(U_{\overline{\mathcal{L}_1} })$ and
		$\psi \in\ker(U-\lambda) \setminus \{0\}$, 
		$f:= d\psi \in \ker(T^2-1)^{\perp}$ and $\psi$ 
		satisfies the following equation:
		\begin{align}
			\label{eq:psiKlambdaf}
			\psi=\frac{a-b}{\lambda^2-b^2}K_{\lambda}f.
		\end{align}
		In particular, $\psi$ is in $\mathcal{L}_1.$ 
	\end{lemma}
	
	\begin{proof}
		From Proposition \ref{Tf=phif}, we have $Tf=\varphi_{a,b}(\lambda)f$. Moreover, since $|\varphi_{a,b}(\lambda)|<1$ holds from Remark \ref{abNotinSigmaUL1}, we obtain $f\in\ker(T^{2}-1)^{\perp}$. 
		It is shown that  $\psi\in\mathcal{L}_1$ if \eqref{eq:psiKlambdaf} holds.
		
		Next, we prove \eqref{eq:psiKlambdaf}. 
		Using the arguments in the proof of \cite[Proposition 5.1]{SS.19}, the vector $\psi_0 :=\psi-d^\ast f$ is represented by 
		\begin{align*}
			\psi_0=(\lambda-bS)^{-1}(aSd^\ast - \lambda d^\ast)f.
		\end{align*}
		By the direct calculation, we see that $(\lambda-bS)^{-1}=(\lambda+bS)(\lambda^2-b^2)^{-1}$. Therefore, 
		\begin{align*}
			\psi
			&=d^\ast f+\psi_0\\
			%&=\bigl(1+(\lambda-bS)^{-1}(aS-\lambda)\bigr)d^\ast f\\
			%&=\frac{1}{\lambda^2-b^2}\bigl(\lambda^2-b^2+(\lambda+bS)(aS-\lambda)\bigr)d^\ast f\\
			%&=\frac{a-b}{\lambda^2-b^2}(b+\lambda S)d^\ast f\\
			&=\frac{a-b}{\lambda^2-b^2}K_{\lambda}d\psi.
		\end{align*}
		Hence we get the desired result.
	\end{proof}
	
	{\it Proof of} $\eqref{eq:sigmapUL1=phiinverse:dimension}$\\
	Recall $K_{\lambda}\ker(T-\varphi_{a,b}(\lambda)) \subset \ker(U-\lambda)$ for $\lambda \in \varphi_{a,b}^{-1}(\sigma_\mathrm{p}(T) \setminus \{\pm1\})$. 
	Since $K_\lambda$ is injective if $\lambda\in\sigma_\mathrm{p}(U_{\overline{\mathcal L_1}})$, the above lemma provide us that $K_{\lambda}$ is bijection from  $\ker(T-\varphi_{a,b}(\lambda))$ onto $\ker(U-\lambda)$. 
	Hence, we obtain \eqref{eq:sigmapUL1=phiinverse:dimension}. $\ \hfill\Box$\\ 
	Summarizing \eqref{eq:decomposeL0andLperp}, 
	Proposition \ref{pointspectrum} and \ref{pointUL1},
	we get Theorem \ref{Maineigenvalue}.

	%%%%%%%%%%%%%%%%%%%%%%%%%%%%%%%%%%%%%%%%%%%%%%%%%%%%%%%%%%%%%%%%%%%%%%%%
	\section{Continuous spectrum of time evolution}
	%%%%%%%%%%%%%%%%%%%%%%%%%%%%%%%%%%%%%%%%%%%%%%%%%%%%%%%%%%%%%%%%%%%%%%%%
	In this section, we will prove the next theorem:
	%%% Statement %%%
	\begin{thm}\label{MainConti}
		The following assertions hold:
		\begin{align}
			\label{eq:MainConti1}
			&\sigma_\mathrm{c}(U) \setminus \{\pm a,\pm b\}
			= \varphi_{a,b}^{-1}(\sigma_\mathrm{c}(T)\setminus\{\pm 1\} ).
		\end{align}
		Moreover, the following hold:
		\begin{enumerate}[{\rm (i)}]
			\item $\pm a\in\sigma_\mathrm{c}(U)$ if and only if $\pm1\in\sigma_\mathrm{c}(T)$.
			\item If $\mp b \in \sigma_\mathrm{c}(U)$, then $\pm1\in\sigma(T)$.
			\item If $\pm 1 \in \sigma_\mathrm{c}(T)$, then $\mp b\in\sigma(U)$.
		\end{enumerate}
	\end{thm}

	To prove this theorem, we introduce some notations. The symbols 
	\begin{align*}
		\sigma_{\rm ap}(A)&:=\left\{z\in \mathbb{C} \ \middle| \ 
		\begin{array}{l} 
			\mbox{there exists $\{\psi_n\}_{n=1}^{\infty}\subset \mathcal{N}$ such that}\\ \mbox{$\|\psi_n\|=1$ and $\lim_{n \rightarrow \infty} \|(A-z)\psi_n\| =0$}
		\end{array}
		\right\},
	\end{align*}
	$\sigma_\mathrm{r}(A)$ and $\rho(A)$ are called respectively the approximate point spectrum, residual spectrum and resolvent set of an operator $A$ on a Hilbert space $\mathcal{N}$. 
	In general, the following relation holds (see \cite[Theorem 3.1.20]{HY}):
	\begin{align*}
		\sigma_\mathrm{p}(A)\cup \sigma_\mathrm{c}(A) \subset \sigma_{\rm ap}(A) \subset \sigma(A).
	\end{align*}

	\begin{rem}
		In general, for the directsum operator $A\oplus B$ of linear operators $A$ and $B$ which respectively act on Hilbert spaces $\mathcal{H}_1$ and $\mathcal{H}_2$, we have 
		\begin{align}
			\label{eq:sigmarDirectsum_general}
			\sigma_\mathrm{r}(A\oplus B) &= 
			\bigl(\sigma_\mathrm{r}(A) \cap \sigma_\mathrm{r}(B)\bigr) \cup
			\bigl(\sigma_\mathrm{r}(A) \cap \sigma_\mathrm{c}(B)\bigr) \cup
			\bigl(\sigma_\mathrm{r}(A) \cap \rho(B)\bigr)\\
			& \quad \cup
			\bigl(\sigma_\mathrm{c}(A) \cap \sigma_\mathrm{r}(B)\bigr) \cup
			\bigl(\rho(A) \cap \sigma_\mathrm{r}(B)\bigr),\nonumber\\
			\label{eq:sigmacDirectsum_general}
			\sigma_\mathrm{c}(A\oplus B) &= 
			\bigl(\sigma_\mathrm{c}(A) \cap \sigma_\mathrm{c}(B)\bigr) \cup
			\bigl(\sigma_\mathrm{c}(A) \cap \rho(B)\bigr) \cup
			\bigl(\rho(A) \cap \sigma_\mathrm{c}(B)\bigr)
		\end{align}
		by the definition of spectra. 
		From $\sigma_\mathrm{r}(U_{\mathcal{L}_1^\perp})=\emptyset=\sigma_\mathrm{c}(U_{\mathcal{L}_1^\perp})$,  $\mathcal{L}_1^\perp=\mathcal{L}_0\oplus\mathcal{L}^\perp$ and 
		\eqref{eq:decomposeL0andLperp}
		we have
		\begin{align}
			\label{eq:sigmarDirectsum}
			\sigma_\mathrm{r}(U) &= 
			\sigma_\mathrm{r}(U_{\overline{\mathcal{L}_1}}) \setminus 
			\bigl(\{a\}^{m_{+} }\cup\{-a\}^{m_{-} }\cup\{-b\}^{M_{+} }\cup\{b\}^{M_{-} }\bigr),\\
			\label{eq:sigmacDirectsum}
			\sigma_\mathrm{c}(U) &= 
			\sigma_\mathrm{c}(U_{\overline{\mathcal{L}_1}}) \setminus 
			\bigl(\{a\}^{m_{+} }\cup\{-a\}^{m_{-} }\cup\{-b\}^{M_{+} }\cup\{b\}^{M_{-} }\bigr)
		\end{align}
		as $A=U_{\overline{\mathcal{L}_1}}$ and $B=U_{\mathcal{L}_1^\perp}$. 
		This and the absence of $\sigma_\mathrm{r}(U)$ give 
		$\sigma_\mathrm{r}(U_{\overline{\mathcal{L}_1}})\setminus\{\pm a, \pm b\} =\emptyset$ and hence we get 
		\begin{align}\label{eq:sigmaL1_subset_sigmapc}
			\sigma(U_{\overline{\mathcal{L}_1}})\setminus\{\pm a, \pm b\}
			\subset \sigma_\mathrm{p}(U_{\overline{\mathcal{L}_1}}) \cup \sigma_\mathrm{c}(U_{\overline{\mathcal{L}_1}}).
		\end{align}
		In particular, if $m_{\pm}=0$, then we obtain   
		\begin{align}\label{eq:sigmaL1anda_subset_sigmapc}
			\sigma(U_{\overline{\mathcal{L}_1}})\setminus \{\pm b\}
			\subset \sigma_\mathrm{p}(U_{\overline{\mathcal{L}_1}}) \cup \sigma_\mathrm{c}(U_{\overline{\mathcal{L}_1}}).
		\end{align}
	\end{rem}

	%spectral property 
	\begin{comment}
	\begin{prop}\label{spectra_chiral}
	For a bounded operator $W$ and unitary involution $\Gamma$ on $\mathcal{H}$, 
	assume that $\Gamma W \Gamma^\ast = W^\ast$ holds. 
	Then, the residual spectrum of $U$ is empty. 
	\end{prop}
	
	\begin{proof}
	Generally, for an linear operator $A$ and a unitary operator $\eta$ on $\mathcal{H}$, 
	the following equation holds:
	\begin{align}
	\sigma_{\sharp}(\eta A\eta^\ast) = \sigma_{\sharp}(A)
	\end{align}
	for each $\sharp = \mathrm{p, c, r}$. 
	This and the assumption imply 
	\begin{align}\label{sigmaW=sigmaW*}
	\sigma_{\sharp}(W) = \sigma_{\sharp}(W^\ast)
	\end{align}
	By \cite[Proposition 3.1.19 (3)]{HY}, 
	$\sigma_\mathrm{p}(W^\ast)$ includes
	the adjoint of $\sigma_\mathrm{r}(W)$. 
	Now \eqref{sigmaW=sigmaW*} and the definition of the spectrum provide us  
	$\sigma_\mathrm{r}(W^\ast) = \sigma_\mathrm{r}(W)^\ast$ and 
	$\sigma_\mathrm{r}(W^\ast) \cap \sigma_\mathrm{p}(W^\ast)=\emptyset$
	respectively, 
	since the residual spectrum of $W$ is empty.
	\end{proof}
	This proposition give us that the residual spectrum of $U$ is empty.
	\end{comment}
	\begin{comment}
	We define the following sets 
	\begin{align*}
	K_1:=\sigma_\mathrm{p}(T)\cap\{\pm1\},\quad 
	K_2:=\sigma_\mathrm{p}(U)\cap\{\pm a, \pm b\}.
	\end{align*}
	Recall that $K_2=\sigma_\mathrm{p}(U_{\overline{\mathcal{L}_0} })\cup \sigma_\mathrm{p}(U_{\mathcal{L}^{\perp} })$ holds. 
	\end{comment}
	
	To prove Theorem \ref{MainConti}, we need the following lemmas.
	\begin{lemma}\label{sigmacL1_subset_sigmacT}
		The following holds:
		\begin{align}
			\label{eq:contispeL1}
			%\sigma(U_{\overline{\mathcal{L}_1} })\setminus (K_2\cup \{\pm b\})
			%\subset \varphi_{a,b}^{-1}(\sigma(T)). \\
			\sigma_\mathrm{c}(U_{\overline{\mathcal{L}_1} })\setminus\{\pm a, \pm b\}
			\subset
			\varphi_{a,b}^{-1}(\sigma_\mathrm{c}(T)). 
			%\left\{\lambda \in \mathbb{C} \ \middle| \ \frac{\lambda-ab\lambda^{-1} }{a-b} \in \sigma(T) \right\}
		\end{align}
	\end{lemma}
	
	\begin{proof}
		Let $\lambda\in \sigma_\mathrm{c}(U_{\overline{\mathcal{L}_1}})$. 
		\begin{comment}
		From Proposition \ref{res_sp_U}, 
		we obtain 
		$\sigma(U_{\overline{\mathcal{L}_1}}) = 
		\sigma_\mathrm{ap}(U_{\overline{\mathcal{L}_1}})$. 
		\end{comment}
		There exists a sequence 
		$\{\psi_n\}_{n=1}^{\infty}\subset \overline{\mathcal{L}_1}$ such that 
		$\|\psi_n\|=1$ for all $n\in \mathbb{N}$ and 
		$\lim_{n \rightarrow \infty}\|(U-\lambda)\psi_n\|=0$.
		For $f_n :=d\psi_n$ and $\theta_n:=\psi_n - d^\ast f_n$ 
		the fact that $d^\ast$ is isometric provides us, for all $n\in\mathbb{N}$, 
		\begin{align}
			\label{eq:psiftheta}
			\|\psi_n\|^2=\|\theta_n\|^2+\|f_n\|^2.
		\end{align}
		Assume that $f_n$ tends to zero as $n\rightarrow \infty,$ 
		then $\lim_{n \rightarrow \infty}\|\theta_n\|=1$. 
		In particular, we can choose $n_0\in\mathbb{N}$ satisfying 
		$\inf_{n\geq n_0}\|\theta_n\|>0$. 
		Since 
		\begin{comment}
		$\lim_{n \rightarrow \infty}f_n = 0$ and 
		\end{comment}
		$d^\ast $ is bounded, we obtain 
		\begin{align*}
			U\psi_n&=SC(\theta_n + d^\ast f_n)\\
			%&=bS\theta_n +aSd^\ast f_n\\
			&=bS\theta_n +o(1) \quad (n\to \infty). 
		\end{align*}
		Thus, it holds that 
		\begin{align*}
			(U-\lambda)\psi_n=b\left(S-\frac{\lambda}{b}\right)\theta_n + o(1)
			\quad (n\to \infty).
		\end{align*}
		This implies that $(S-\lambda/b)\theta_n$ tends to zero as $n\rightarrow \infty$ 
		because so is $(U-\lambda)\psi_n$. 
		This fact and $\inf_{n\geq n_0}\|\theta_n\|>0$ 
		give us $\lim_{n\rightarrow \infty}\|(S-\lambda/b)\tilde{\theta}_n \|=0$ 
		where $\tilde{\theta}_n:=\theta_n/\|\theta_n\|$ for $n\geq n_0.$ 
		Hence $\lambda/b$ is in $\sigma(S).$ 
		Since the spectrum of $S$ consists of $\pm1$, we have $\lambda\in\{\pm  b\}.$
		
		We assume that $\lambda\neq \pm b$. From the above argument, we see that $f_n$ does not converge to zero. 
		In this case, we can show that $\varphi_{a,b}(\lambda) \in \sigma(T)$ holds from the similar way of the proof in 
		\cite[Lemma 6.2. (1)]{SS.19}. 
		In the case of $\lambda\neq\pm a, \pm b$, if $\varphi_{a,b}(\lambda) \in \sigma_\mathrm{p}(T)$, then it is contradict to \eqref{eq:sigmapUL1=phiinverse} since $\varphi_{a,b}(\lambda)\neq\pm1$. 
		Hence, we obtain $\lambda \in \varphi_{a,b}^{-1}(\sigma_\mathrm{c}(T))$ for any $\lambda \in \sigma_\mathrm{c}(U_{\overline{\mathcal{L}_1} })\setminus\{\pm a, \pm b\}$.
	\end{proof}

	\begin{lemma}\label{sigmacT_subset_sigmacL1}
		The following assertion holds:
		%$\varphi_{a,b}^{-1}(\sigma(T))\setminus K_2\subset\sigma(U_{\overline{\mathcal{L}_1} })$.
		\begin{equation*}
			\varphi_{a,b}^{-1}(\sigma_\mathrm{c}(T))\setminus\{\pm b\}\subset
			\sigma_\mathrm{c}(U_{\overline{\mathcal{L}_1} }).
		\end{equation*}
	\end{lemma}

	\begin{proof}
		Let $\lambda\in\varphi_{a,b}^{-1}(\sigma_\mathrm{c}(T))\setminus\{\pm b\}$. 
		Since $\varphi_{a,b}(\lambda)\in\sigma_\mathrm{c}(T)$, there exists a sequence $\{f_n\}\subset\mathcal{K}$ such that $\|f_n\|=1$ for all $n\in\mathbb{N}$ and
		\[\lim_{n\to\infty}\|\bigl(T-\varphi_{a,b}(\lambda)\bigr)f_n\|=0.\]
		Then, we set $\psi_n:=(b+\lambda S)d^\ast f_n$. 
		It is easy to see that $\psi_n\in\mathcal{L}_1$ and 
		\begin{align*}
			\|\psi_n\|^2 
			&=|b|^2+2b\Re (\lambda\langle f_n,Tf_n\rangle) + |\lambda|^2\\
			%&=|b|^2+2b\Re (\lambda\langle f_n,\varphi_{a,b}(\lambda)f_n\rangle)
			%+ |\lambda|^2 + 2b\Re \left(\lambda\langle f_n, \bigl(T-\varphi_{a,b}(\lambda)\bigr)f_n\rangle\right)\\
			%&\geq |b|^2-2|b||\lambda| |\varphi_{a,b}(\lambda)|+|\lambda|^2-2|b||\lambda|\|\bigl(T-\varphi_{a,b}(\lambda)\bigr)f_n\|\\
			%&\geq |b|^2-2|b||\lambda|+|\lambda|^2-2|b||\lambda|\|\bigl(T-\varphi_{a,b}(\lambda)\bigr)f_n\|\\
			%&=
			&\geq 
			\bigl(|b|-|\lambda|\bigr)^2-2|b||\lambda|\|\bigl(T-\varphi_{a,b}(\lambda)\bigr)f_n\|
		\end{align*}
		holds from $|\varphi_{a,b}(\lambda)|\leq 1$.
		Recall $|\lambda|=\sqrt{-ab}$ when 
		$ab<0$ and $\lambda\in\mathbb{C}\setminus\mathbb{R}$, 
		by $\lambda\neq\pm b$ we have 
		\[\liminf_{n\to\infty}\|\psi_n\|^2\geq(|b| - |\lambda|)^2>0.\]
		Therefore we can choose a subsequence $\{\phi_n\}\subset\{\psi_n\}$ which satisfies $\inf_n\|\phi_n\|>0$. Then,
		\begin{align*}
			U\phi_n
			&=SC(b+\lambda S)d^\ast f_n\\
			%&=abSd^\ast f_n+a\lambda Sd^\ast Tf_n+b\lambda S(1-d^\ast d)Sd^\ast f_n\\
			%&=abSd^\ast f_n+(a-b)\lambda Sd^\ast Tf_n+b\lambda d^\ast f_n\\
			%&=abSd^\ast f_n+(a-b)\lambda\varphi_{a,b}(\lambda)Sd^\ast f_n+b\lambda d^\ast f_n+(a-b)\lambda Sd^\ast\bigl(T-\varphi_{a,b}(\lambda)\bigr)f_n\\
			%&=\Bigl(b\lambda+\bigl((a-b)\lambda\varphi_{a,b}(\lambda)+ab\bigr)S\Bigr)d^\ast f_n+(a-b)\lambda Sd^\ast\bigl(T-\varphi_{a,b}(\lambda)\bigr)f_n\\
			%&=(b\lambda+\lambda^2S)d^\ast f_n+(a-b)\lambda Sd^\ast\bigl(T-\varphi_{a,b}(\lambda)\bigr)f_n\\
			&=\lambda\phi_n+(a-b)\lambda Sd^\ast \bigl(T-\varphi_{a,b}(\lambda)\bigr)f_n.
		\end{align*}
		Thus, $\lim_{n\to\infty}\|(U-\lambda)\phi_n\|=0$ holds and hence 
		$\lambda\in\sigma(U_{\overline{\mathcal L_1}})$. 
		Furthermore, if $\lambda\neq\pm a$,  then \eqref{eq:sigmapUL1=phiinverse} and \eqref{eq:sigmaL1_subset_sigmapc}
		imply $\lambda\in\sigma_\mathrm{c}(U_{\overline{\mathcal L_1}})$.  
		If $\lambda=\pm a$, this indicates $\pm1\in\sigma_\mathrm{c}(T)$, in particular $m_\pm=0$. 
		Recall Remark \ref{abNotinSigmaUL1}
		and hence $\lambda\in\sigma_\mathrm{c}(U_{\overline{\mathcal L_1}})$
		holds by \eqref{eq:sigmaL1anda_subset_sigmapc}.
		Now we obtain our assertion.
	\end{proof}

	{\it Proof of Theorem} $\ref{MainConti}$\\
	Let us first prove \eqref{eq:MainConti1}. 
	Lemmas \ref{sigmacL1_subset_sigmacT} and \ref{sigmacT_subset_sigmacL1} provide 
	\begin{align*}
		\sigma_\mathrm{c}(U_{\overline{\mathcal{L}_1} })\setminus\{\pm a,\pm b\} 
		&= \varphi_{a,b}^{-1}(\sigma_\mathrm{c}(T))\setminus\{\pm a, \pm b\}\\
		&= \varphi_{a,b}^{-1}(\sigma_\mathrm{c}(T)\setminus \{\pm1\}).
	\end{align*}
	Thus \eqref{eq:sigmacDirectsum} provides
	\begin{equation*}
		\sigma_\mathrm{c}(U) \setminus \{\pm a,\pm b\}
		= \varphi_{a,b}^{-1}(\sigma_\mathrm{c}(T)\setminus \{\pm1\}).
	\end{equation*}
	If $\pm1\in\sigma_\mathrm{c}(T)$, then $\pm a \in \sigma_\mathrm{c}(U)$ holds from Lemma\ref{sigmacT_subset_sigmacL1} and $\mp b\in\sigma(U)$ holds by a similar argument of \cite[Proof of Lemma 6.2]{SS.19}. 
	Let $\nu=\pm a,\mp b$. 
	If $\nu \in \sigma_\mathrm{c}(U)$, then \eqref{eq:sigmacDirectsum} gives $N_\nu=0$,where 
	\begin{align*}
		N_\nu=
		\begin{cases}
			m_+, & \nu=a,\\
			m_-, & \nu=-a,\\
			M_+, & \nu=-b,\\
			M_-, & \nu=b,
		\end{cases}
	\end{align*}
	and $\nu \in \sigma_\mathrm{c}(U_{\overline{\mathcal{L}_1} })$.
	Hence discussion of the accumulation point implies $\pm1 \in \sigma(T)$. 
	In particular, for $\nu=\pm a$, $m_{\pm}=0$ holds and that gives us $\pm1\not\in\sigma_{\mathrm{p}}(T)$. 
	Hence $\pm1$ is in $\sigma_\mathrm{c}(T)$ respectively.
	$\hfill \Box$

	Theorem \ref{Main} is a combination of 
	Theorem \ref{Maineigenvalue} and \ref{MainConti}.
	
	\begin{rem}
		In this subsection, we use the approximate point spectrum to prove our statement. 
		This approach is derived from the absence of the residual spectrum of  $U$ and that is obtained by chiral symmetry and a reflection symmetry of eigenvalues about the real axis. 
	\end{rem}

	%%%%%%%%%%%%%%%%%%%%%%%%%%%%%%%%%%%%%%%%%%%%%%%%%%%%%%%%%%%%%%%%%%%%%%%%%%%%%
	\section{Applications}
	%%%%%%%%%%%%%%%%%%%%%%%%%%%%%%%%%%%%%%%%%%%%%%%%%%%%%%%%%%%%%%%%%%%%%%%%%%%%%
	In this section, we introduce two applications. In the first, we consider Mochizuki-Kim-Obuse (MKO) model which is a non-unitary quantum walk researched by \cite{MKO.16}. 
	Next, we consider the application to the Ihara zeta function.
	\subsection{Mochizuki-Kim-Obuse model}
	Let $\gamma >0, \psi ,\theta _j\in [0,2\pi)$ for $j=1,2.$ We define some matrices $G, \Phi $ and $\tilde{C}(\theta _j)$ on $\mathbb{C}^2$ as 
	\begin{align*}
		G:=
		\begin{pmatrix}
			e^\gamma &0\\
			0&e^{-\gamma}
		\end{pmatrix},
		\quad \Phi :=
		\begin{pmatrix}
			e^{i\phi }&0\\
			0&e^{-i\phi }
		\end{pmatrix},
		\quad \tilde{C}(\theta _j)=
		\begin{pmatrix}
			\cos \theta_j&i\sin \theta_ j\\
			i\sin \theta_j&\cos \theta _j
		\end{pmatrix}
		,\quad j=1,2.
	\end{align*}
	We write the multiplication operator on $\ell^2(\mathbb{Z};\mathbb{C}^2)$ by $G, \Phi $ and $\tilde{C}(\theta _j)$ by the same symbols.
	For $\Psi \in \ell^2(\mathbb{Z};\mathbb{C}^2),$ the operator $\tilde{S}$ is defined by 
	\begin{align*}
		(\tilde{S}\Psi )(x):=
		\begin{pmatrix}
			\Psi _1(x+1)\\
			\Psi _2(x-1)
		\end{pmatrix}
		,\quad \Psi (x)=
		\begin{pmatrix}
			\Psi _1(x)\\
			\Psi _2(x)
		\end{pmatrix}
		,\quad x\in \mathbb{Z}.
	\end{align*}
	The MKO model is defined $U_{\gamma }$ on $\ell^2(\mathbb{Z};\mathbb{C}^2)$ by
	\begin{align*}
		U_{\gamma }:=\tilde{S}G\Phi \tilde{C}(\theta _2)\tilde{S}G^{-1}\Phi \tilde{C}(\theta _1),\quad \theta _1,\theta _2\in [0,2\pi).
	\end{align*} 
	In \cite{MKO.16}, $\Phi$ is called the Phase operator. 
	For $\gamma >0$, $e^{-\gamma}$ and $e^{\gamma}$ mean the loss and gain of the photon. 
	If the parameter $\gamma$ is large enough, then the loss and gain amplify. 
	Thus $G$ is called the loss and gain operator in \cite{MKO.16}. 
	In the case of $\gamma =0$ or $\sin \theta _2=0$, $U_{\gamma}$ is unitary. 
	However if $\gamma >0$ and $\sin \theta _2\not =0$, then $U_{\gamma }$ is not unitary. 
	Thus $U_{\gamma }$ is a time evolution operator of a non-unitary quantum walk. 
	Moreover, we remark that $U_{\gamma}$ is not normal. Thus we can not use some general theories for the spectral analysis. 
	In \cite{MKO.16}, the parameters $\theta _1$ and $\theta _2$ may depend on $x\in \mathbb{Z}$. 
	However, we take $\theta _1$ and $\theta _2$ such that these do not depend on $x\in \mathbb{Z}$ in this paper. By the next lemma, the spectrum mapping theorem is applicated for the MKO model.
	
	\begin{lemma}\cite[Theorem B]{AsaharaEtAl}\label{lem:app}
		There exists a unitary operator $\eta $, a self-adjoint unitary operator $S_{\rm mko}$ and a self-adjoint operator $C_{\rm mko}$ on $\ell^2(\mathbb{Z}; \mathbb{C}^2)$ such that 
		$\eta^\ast U_{\gamma }\eta =S_{\rm mko}C_{\rm mko}.$
	\end{lemma}
	\begin{proof}
		We set $\eta :=\sigma _2\tilde{C}(\theta_1)\tilde{S}\sigma _2$ where $\sigma _2$ is the second Pauli matrix i.e., 
		$\sigma _2=
		\begin{pmatrix}
			0&-i\\
			i&0
		\end{pmatrix}$. 
		Then $\eta $ is a unitary operator on $\mathcal{H}$ since $\sigma _2\tilde{C}(\theta _1)$ and $\tilde{S}\sigma _2$ are unitary operators. 
		We define $S_{\rm mko}$ and $C_{\rm mko}$ as $S_{\rm mko}:=\tilde{S}\tilde{C}(\theta _1)\tilde{S}\sigma _2$ and $C_{\rm mko}:=\sigma _2G\Phi\tilde{C}(\theta _2)G^{-1}\Phi$. 
		Then $S_{\rm mko}$ is unitary self-adjoint and $C_{\rm mko}$ has only two real eigenvalues since $\sigma _2\tilde{C}(\theta _1)$ and $\tilde{S}\sigma _2$ are unitary self-adjoint and $C_{\rm mko}$ is a $2\times 2$ Hermitian matrix. 
		Moreover, we obtain $\eta^\ast U_{\gamma}\eta =S_{\rm mko}C_{\rm mko}.$
	\end{proof}
	Lemma~\ref{lem:app} means that the spectra of $U_{\gamma}$ and $S_{\rm mko}C_{\rm mko}$ are equal. As a matter of fact, the spectrum of $U_{\gamma }$ is included in the unit circle and the real line.
	In what follows, we consider the case of $\sin \theta _1\sin \theta _2>0.$ In other cases, we can obtain the results by the same arguments.
	Let $\gamma _0,\ \gamma _1>0$ be 
	\begin{align*}
		\gamma_i:=\frac{1}{2}\log\left( \Gamma_i + \sqrt{\Gamma_i^2-1 } \right),\ \ \Gamma_i=-\frac{1}{ap}\left( 1+(-1)^{i+1}|qb|\right),\ i=0,1
	\end{align*}
	where $p:=-\sin \theta_1,\ q:=-i\cos(\theta_1),\ a:=\sin \theta_2,\ b:=-ie^{-2i\phi }\cos \theta_2.$ Then, by the direct calculation, we obtain the next proposition.
	
	\begin{prop}{\rm \cite[Theorem C]{AsaharaEtAl}}\label{MKOspec}
		For any $\gamma >0$, the following equations hold:
		\begin{align*} 
			\sigma(U_{\gamma })=\sigma_\mathrm{c}(U_{\gamma }), \quad
			\sigma_\mathrm{p} (U_\gamma) = \sigma_\mathrm{r} (U_\gamma) = \emptyset.
		\end{align*}
		Moreover, the spectrum of $U_{\gamma }$ holds that 
		\begin{align*}
			\sigma (U_{\gamma })=
			\begin{cases}
				\left\{ e^{i\xi}\ \Big|\ \cos \xi\in [m_{\gamma}, M_{\gamma}] \right\}, & 0<\gamma \leq\gamma _0,\vspace{3pt}\\
				\left\{ e^{i\xi }\ \Big|\  \cos \xi\in [-1, M_{\gamma}] \right\}\cup 
				[f_- (m_{\gamma }), f_+ (m_{\gamma})], 
				& \gamma _0< \gamma < \gamma _1,\vspace{3pt}\\
				[f_-(m_{\gamma }), f_-(M_{\gamma })]
				\cup[f_+(M_{\gamma }), f_+(m_{\gamma })], 
				& \gamma _1\leq\gamma 
			\end{cases}
		\end{align*}
		where
		\begin{align*}
			f_{\pm}(x)=x\pm\sqrt{x^2-1},\  
			m_{\gamma}:=ap\cosh(2\gamma )-|qb|,\ M_{\gamma}:=ap\cosh(2\gamma )+|qb|.
		\end{align*}
	\end{prop}
	By Proposition \ref{MKOspec}, $\sigma (U_{\gamma})$ is included in the unit circle and the real line. Moreover, the spectrum of $U_{\gamma}$ varies greatly with the parameter $\gamma >0.$ The main theorem is hoped to the applications to the spectral analysis of the MKO model in the spatial dependence case. 
	
	\subsection{Ihara zeta function}
	Originally, the Ihara zeta function was 
	defined by Y.Ihara in the context of discrete subgroups of the $p$-adic special linear group \cite{Ihara.66}.
	In this paper we introduce the definition of the Ihara zeta function in the graph theoretical setting by \cite{Sunada.89}. 
	Let $G=(V,E)$ be a connected, finite and $k$-regular graph, that is, $|V|<\infty$ and $\mathrm{deg}(u)<\infty$. 
	Then the Ihara zeta function can be defined by the following Euler product as an analogue of the Riemann zeta function: 
	\[ \zeta_G(u)=\prod_{[C]}\left(1-u^{|C|}\right)^{-1}, \]
	where $[C]$ runs over all equivalence classes of prime and reduced cycles of $G$ and $|C|$ is the length of the cycle. 
	Let $A=A(G)$ be the set of symmetric arcs. 
	Here if $\{u,v\}\in E$, 
	then the induced symmetric arcs are $(u,v)$ and $(v,u)$ which represent the arcs from the origin vertex $u$ to the terminal vertex $v$ and vice versa, respectively.
	For any countable set $\Omega$, let $\mathbb{C}^\Omega$ be the linear space whose standard basis are labeled by the elements of $\Omega$ 
	and inner product is standard. 
	Bass and Hashimoto \cite{Bass.92, Hashimoto.89} showed the following determinant expression of the Ihara zeta function. 
	\[ 1/\zeta_G(u)=\det(1_{\mathbb{C}^A}-u(B'-J)), \]
	where $B'$ and $J$ are operators on $\mathbb{C}^A$ such that 
	$(B'\psi)(e)=\sum_{e'\in A\;:\;o(e)=t(e')}\psi(e')$ and 
	\begin{align}
		(J\psi)(e)=\psi(\bar{e})\label{eq:J}
	\end{align}
	for any $\psi\in \mathbb{C}^A$. 
	Remark that $(B')_{e,e'}=1$ if and only if $e'$ meets $e$, 
	and the operation $J$ corresponds to the back tracking. 
	Thus $(B'-J)_{e,e'}=1$ if and only if $e'$ meets $e$ but $e'\neq \bar{e}$. 
	
	Now let us take further deformation of this expression connecting to quantum walks. 
	Let $U$ be the time evolution operator of the Grover walk on $G$ for $k>2$ such that
	\[ (U\psi)(e)=-\psi(\bar{e})+\frac{2}{k}\sum_{t(e')=o(e)}\psi(e') \] 
	for any $\psi\in \mathbb{C}^A$ and $e\in A$.
	The positive support of $U$ is defined by 
	\[ (U^+)_{e,e'}=\begin{cases} 1, & \text{$(U)_{e,e'}>0$,}\\ 0, & \text{$(U)_{e,e'}\leq 0$. } \end{cases} \]
	Then interestingly, we can see that $U^+$ coincides with $B'-J$~\cite{RenEtAl.11} because the reflection amplitude of Grover walk corresponding to the back tracking 
	is $2/k-1<0$ and the transmitting amplitude is $2/k>0$.
	Thus, $U^+ = B'-J$
	and
	\begin{align*}
		1/\zeta_G(u)=\det(1_{\mathbb{C}^A}-uU^{+}). 
	\end{align*}
	An attractive study direction from this fact is an investigation of the extension of this zeta function 
	by $1/\zeta_G^{(n)}(u):=\det(1_{\mathbb{C}^A}-u(U^n)^+)$ 
	concerning the graph isomorphic problem see e.g.,\cite{EmmsEtAl.06,GG.11,KS.12,HKSS.14,KSS.19}. 
	\begin{thm}
		Let $G=(V,E)$ be set as the above. 
		Let $M$ be the adjacency matrix of $G$; that is, $(Mf)(u)=\sum_{t(e)=u}f(o(e))$ for any $f\in \mathbb{C}^V$ and $u\in V$. Then we have
		\begin{align}
			&\sigma(U^+)=\left\{\lambda=\frac{\mu}{2}\pm \frac{1}{2}\sqrt{\mu^2-4(k-1)} \ 
			\middle| \ \mu\in\sigma(M)\right\}
			\cup \{\pm1\}^{M_{\pm}} \cup \{\pm(k-1)\}^{m_{\pm}} \\
			&\dim \ker(\lambda - U^+) = \dim (\mu - M)
		\end{align}
	\end{thm}
	This theorem is well-known conclusion, but we give another proof. 
	To use the Theorem \ref{Main}, we show the next lemma.
	\begin{lemma}
		There exist a unitary self-adjoint operator $S$ on $\mathbb{C}^{A}$ and coisometry $d:\mathbb{C}^{A}\to \mathbb{C}^V$ satisfying 
		\[ U^+=S(kd^\ast d-I_{\mathbb{C}^A}). \]
	\end{lemma}
	To this end, we define $K_{\rm in}$ and $K_{\rm out}$ as
	letting $K_{in}\colon \mathbb{C}^{A}\to \mathbb{C}^V$ be an incidence matrix with respect to a terminal vertex of arcs such that 
	$(K_{in}\psi)(u)=\sum_{t(e)=u}\psi(e)$. The adjoint is expressed by $(K_{in}^\ast f)(e)=f(t(e))$. 
	In the same way, let $K_{out}\colon \mathbb{C}^{A}\to \mathbb{C}^V$ be an incidence matrix with respect to origin vertex of arcs such that 
	$(K_{out}\psi)(u)=\sum_{o(e)=u}\psi(e)$. The adjoint is expressed by $(K_{out}^\ast f)(e)=f(o(e))$.
	Then it is easy to see that 
	\[ B'=K_{out}^\ast K_{in},\;M=K_{in}K_{out}^\ast, \]
	where $M$ is the adjacency matrix of $G$. 
	Noting $K_{out}=K_{in}J$, and setting $d$ by $d=1/\sqrt{k} K_{in}$ 
	and $S=J$
	, we have 
	\begin{align*}
		U^+ 
		&= K_{out}^\ast K_{in} - J\\
		&= J(K_{\rm in}^\ast K_{\rm in} - I_{\mathbb{C}^A})\\
		& = S(k d^\ast d - I_{\mathbb{C}^A})
	\end{align*}
	Since $\zeta_G(u)=u^{-|A|}/\det(u^{-1}-U^+)$,  
	the problem is switched to spectral analysis on our quantum walk for $a=k-1$ and $b=-1$ case. 
	By Theorem~3.1 and (3.1), putting $\mu\in \sigma(M)$, we obtain that 
	if $\lambda\in\mathbb{C}$ satisfies $\varphi_{a,b}(\lambda)=(\lambda+(k-1)\lambda^{-1})/k=\mu/k$, then $\lambda\in \sigma(U)$. 
	Such $\lambda$ can be expressed by 
	\[ \lambda=\frac{\mu}{2}\pm \frac{1}{2}\sqrt{\mu^2-4(k-1)}.  \]
	If $|\mu|\leq 2\sqrt{k-1}$, then putting $x(\mu)=\mu/2$ and $y(\mu)=\sqrt{k-1-(\mu/2)^2}$, we have 
	\[ x^2(\mu)+y^2(\mu)=k-1. \]
	Therefore the support of $\lambda\in \sigma(U)$ is included in the circle whose center is $0$ and the radius is $\sqrt{k-1}$ 
	and $\{\mu/2\pm 1/2 \cdot \sqrt{\mu^2-4(k-1)} \;|\; |\mu|> 2\sqrt{k-1},\;\; \mu\in \sigma(M)\}$ on the real line 
	in the complex plain. 
	The former one is called a non-trivial eigenvalues which is an analogue of the pole of the Riemann zeta function. 
	
	Finally, let us compute $m_\pm$ and $M_\pm$ in this case. 
	Let us see $f_+ \in \ker(1-T)$ implies $d^\ast f_+ \in \ker(k-1-U^+)$ in the following, where $T=(1/k) M$. 
	We can check that if $f_+$ is a constant function, then $f_+\in \ker(1-T)$ in this setting. 
	Then by Perron-Frobenius theorem, $\ker(1-T)=\mathbb{C}\{f_+\}$ because ``$1$" is the largest eigenvalue of the positive matrix $T$. 
	Since $d^\ast f_+$ is also a constant function by definition of $d^\ast$ and $f_+$, we have $Sd^\ast f_+=d^\ast f$. 
	Then $U^+d^\ast f_+=S(kd^\ast d-I_{\mathbb{C}^A})d^\ast f_+=(k-1)Sd^\ast f_+=(k-1)d^\ast f_+$. 
	On the other hand, if $f_-\in \ker(-1-T)$ with $f_-\neq 0$, 
	then $G$ must be a bipartite graph from a general spectral graph theory. 
	Decomposing the vertex set of this bipartite graph into $V=X\sqcup Y$ so that ``$t(e)\in X$, $o(e)\in Y$" or 
	``$o(e)\in X$, $t(e)\in Y$" for any $e\in A$, we obtain that the shape of $f_-$ is 
	\[ f_-(u)=\begin{cases} f_+(u), & \text{$u\in X$,}\\ -f_+(u), & \text{$u\in Y$.} \end{cases}\] 
	Then $Sd^\ast f_-=-d^\ast f$ which 
	implies $U^+d^\ast f_-=S((k-1)dd^\ast-I_{\mathbb{C}^A})d^\ast f_-=Sd^\ast g=-d^\ast f_-$. 
	Therefore, by Theorem~3.1, since $m_\pm=\dim(\ker(\pm 1-T))$, we have 
	\[ m_+=1, \;\; m_-=\begin{cases} 1, & \text{$G$ is bipartite,}\\ 0, & \text{otherwise. } \end{cases} \]
	The value $M_\pm$ can be obtained as $M_\pm=|E|-|V|+m_{\pm}$ (see e.g., \cite{Segawa.15} for the proof). 
	Then we have
	\[ M_\pm=\begin{cases} b_1(G), & \text{$G$ is bipartite,}\\ b_1(G)-1, & \text{$G$ is not bipartite.} \end{cases} \]
	Here $b_1(G):=|E|-|V|+1$ is the first Betti number of $G$. 
	%Next, let us compute $M_\pm$ in this case. 
	%Let $\mathcal{L}:=d^\ast\mathbb{C}^V+d_B^\ast\mathbb{C}^V$. 
	%Then we have 
	%$\dim(\mathcal{L}^\perp)=2|E|-\dim(\mathcal{L})=2|E|-(2|V|-\dim(d^\ast\mathbb{C}^V\cap d_B^\ast\mathbb{C}^V))$. 
	%Remark that $\psi \in d^\ast\mathbb{C}^V\cap d_B^\ast\mathbb{C}^V$, then there exist $f,g\in \mathbb{C}^V$ such that 
	%$\psi=d^\ast f=d_B^\astg$. Since $T=dd_B^\ast=d_Bd^\ast$, we have $f,g\in \ker(1-T^2)$. 
	%Then $d^\ast\mathbb{C}^V\cap d_B^\ast\mathbb{C}^V=d^\ast\ker(1-T^2)$ which implies 
	%\[ \dim(\mathcal{L}^\perp)=2b_1(G)-\bs{1}_{B}(G),  \]
	%where $b_1(G)=|E|-|V|+1$ and $\bs{1}_B(G)=0$ if $G$ is bipartite, $=1$ if $G$ is not bipartite. 
	%Let us see if $\bs{1}_B(G)=0$, then $M_+=\dim(\mathcal{L}^\perp \cap \ker(1+S))$. 
	%Define $\gamma_+_c (a_j)=\gamma_+_c (\bar{a}_j)=(-1)^j$ for a closed cycle $c=(a_1,\dots,a_j)$ with 
	%$t(a_1)=o(a_2)$,\dots, $t(a_{j-1})=o(a_j)$ and $t(a_j)=o(a_1)$. 
	%Remark that since $G$ is bipartite, every closed cycle is an even cycle. 
	%It is easy to check that $\gamma_c\in \ker d$ and $\gamma_c\in \ker(1+S)$. 
	%%
	\subsection{A correlated random walk on a $k$-regular graph}
	Let us consider the following random walk on a connected $k$-regular graph $G=(V,E)$ ($k\geq 2$), which is finite. 
	Let $A$ be the set of the symmetric arcs induced by $E$. 
	We propose the time evolution of the correrated random walk with the parameter $p\in[0,1]$ by 
	\[ (P\psi)(e)=p\psi(\bar{e})+\frac{1-p}{k-1}\sum_{t(e')=o(e), \;e'\neq \bar{e}}\psi(e') \]
	for any $\psi\in \mathbb{C}^A$  and $e\in A$. 
	This means that 
	%the probability that a random walker on an arc $a\in A$ moves to the inverse arc $\bar{a}$ is $p$, while the probability that it moves to the arcs which meats the arc $a$ except its inverse arc $\bar{a}$ is $1-p$ and it chooses uniformly to an arc from the $k-1$ arcs. In other word, 
	the probability that a random walker returns back to the same vertex as one at the previous time is $p$, while the probability that it moves to the other neighbor vertices is $1-p$ and it uniformly chooses a neighbor from the $k-1$ neighbor vertices. 
	Note that if $p=1/k$, then the random walk is reduced to the isotropic random walk. 
	Also note that if the graph is the one-dimensional lattice, then this random walk recovers the model in  \cite{RH, Konno.09}. 
	It is easy to see that $P=SC$ with 
	\[ a=1,\quad b=\frac{pk-1}{k-1}, \]
	$S=J$, where $J$ is defined in \eqref{eq:J}, and the coisometry $d:\mathbb{C}^A \to \mathbb{C}^V$ is 
	\[ (d\psi)(u)=\frac{1}{\sqrt{k}}\sum_{t(e)=u}\psi(e) \]
	for any $\psi\in \mathbb{C}^A$ and $u\in V$. 
	The discriminant operator $T=dSd^*=(1/k)K_{in}JK_{in}^*:=P_0$ is described by the transition matrix of the isotropic random walk on $G$; that is, 
	\[ (P_0f)(u)=\frac{1}{k}\sum_{t(e)=u}f(o(e)) \]
	for any $f\in\mathbb{C}^V$ and $u\in V$. 
	Then from a standard argument on the spectral graph theory, we have 
	\[m_+=1,\quad m_-=\begin{cases} 1, & \text{$G$ is bipartite,}\\ 0, & \text{otherwise,} \end{cases}\]
	\[M_+=b_1(G),\;M_-=\begin{cases} b_1(G), & \text{$G$ is bipartite,}\\ b_1(G)-1, & \text{otherwise.} \end{cases}\]
	By Theorem~\ref{MainConti}, 
	\[\sigma(P)\subset 
	\begin{cases}
		[-1,-r]\cup[r,1], & \text{$1/k\leq p\leq 1$, }\\
		\left\{z\in \mathbb{C} \mid |z|=\sqrt{r}\right\}\cup [-1,-r] \cup [r,1], & \text{$0\leq p<1/k$,}
	\end{cases}\]
	where $r=|pk-1|/(k-1)$. 
	Let us see the transition of $\sigma(P)$ by moving the parameter $p$ from $p=1$ to $p=0$ in the following. 
	If $p=1$, then $\sigma(P)=\{\pm 1\}$; the walk is reduced to just a zigzag walk. As $p$ decreases, the intervals $[-1,-r]\cup [r,1]$ grow up because $r$ is monotone decreasing with respect to $1/k\leq p$. Finally if $p$ reaches to $1/k$, then the two intervals unite. After $p<1/k$, then the gap on the real axis between $(-r,r)$ appears again, moreover some eigenvalues of $P_0$ are inherited to the circle with the radius $\sqrt{r}$ in the complex plane; the circle grows up while the two intervals shrink because $r$ is monotone increasing with respect to $p<1/2$. The final radius of the circle and the final gap size on the real axis at $p=0$ are  $1/\sqrt{k-1}$ and $2/(k-1)$, respectively. 
	On the other hand, Theorem~\ref{MainConti} implies that the geometric information of $b_1(G)$ and the bipartiteness of $G$, which are independent of the parameter $p$, are reflected as the multiplicities $M_{\pm}$ of the eigenvalues $\pm \sqrt{r}$.  
	%%%%%%%%%%%%%%%%%%%%%%%%%%%%%%%%%%%%%%%%%%%%%%%%%%%%%%%%%%%%%%%%%%%%%%%%%%%%%
	\appendix
	\section{Appendix}
	%%%%%%%%%%%%%%%%%%%%%%%%%%%%%%%%%%%%%%%%%%%%%%%%%%%%%%%%%%%%%%%%%%%%%%%%%%%%%
	Here we prove the following proposition used for proving Corollary \ref{cor.2.5} and give two examples of our model, which are one-dimensional non-unitary quantum walks. 
	Let $U$ be defined as in Definition \ref{def.2.3}
	and allow the case $ab=0$. 
	% resolvent of U
	\begin{prop}\label{rho(U)}
		The following statements hold:
		\begin{equation}
			\label{eq:<mina,b}
			\{z\in \mathbb{C} \mid |z|<\min\{|a|, |b|\}  \text{\rm \  or } \max\{|a|, |b|\}<|z|\} \subset \rho(U),%\\
			% \label{eq:<maxa,b}
			% &\{z\in \mathbb{C} \mid |z|>\max\{|a|, |b|\}\} \subset \rho(U). 
		\end{equation}
	\end{prop}
	
	\begin{proof}
		Let $z$ be in the set of the left hand of \eqref{eq:<mina,b}. % or \eqref{eq:<maxa,b}.
		First, we show that $U-z$ is injective. 
		For any $\psi \in \mathcal{H}$ by Schwarz's inequality we have
		\begin{equation*}
			\|(U-z)\psi\|\geq \bigl| \|U\psi\|-|z|\|\psi\| \bigr|.
		\end{equation*}
		From the definition of $C$ and what $d^\ast d$ is a projection, 
		we can show that 
		\begin{align}
			\label{eq:minmax}
			\min\{|a|,|b|\} \|\psi\| \leq \|C\psi\| \leq \max\{|a|,|b|\}\|\psi\|
		\end{align}
		holds. 
		Moreover, we obtain $\|U\psi\|=\|C\psi\|$ by $S^2=I_{\mathcal{H}}$. 
		Thus the inequality  
		\begin{align}
			\label{eq:Uz}
			\|(U-z)\psi\|\geq \bigl|c-|z|\bigr|\|\psi\|
		\end{align}
		holds respectively where $c$ is $\min\{|a|,|b|\}$ or $\max\{|a|,|b|\}$ for each $z$. 
		In both cases $U-z$ is injective. 
		
		Next, we will show that $U-z$ is surjective. Recall the direct sum decomposition of  $\mathcal{H}$ as follows 
		\begin{equation*}
			\mathcal{H}=\ker(U^\ast-\overline{z})\oplus \overline{{\rm ran}(U-z)}.
		\end{equation*}
		Assume that $\phi\in \ker(U^\ast-\overline{z})$, i.e., $CS\phi=U^\ast\phi=\overline{z}\phi.$ Thus we obtain 
		\begin{equation*}
			US\phi=SCS\phi=\overline{z} S\phi.
		\end{equation*}
		This means $S\phi\in\ker(U-\overline{z})$. However, the inequality \eqref{eq:Uz} for $\overline{z}$ 
		provides us $\ker(U-\overline{z})=\{0\}$, since we have $S\phi=0.$ 
		Then, we obtain $\phi=S^2\phi=0$ and this implies that ${\rm ran}(U-z)$ is dense, moreover equals to $\mathcal{H}$ since $U-z$ is bounded from below.
		Hence, $z\in\rho(U)$ because $U-z$ is bijection. 
	\end{proof}
	
	We give two examples of our model.
	First, we give the example of a one-dimensional homogeneous non-unitary quantum walk. 
	\begin{exam}\label{example2}
		For $\phi\in\mathbb{C}^2$ satisfying $\|\phi\|=1$, let 
		$d:\ell^2 (\mathbb{Z};\mathbb{C}^2) \to \ell^2 (\mathbb{Z})$ as 
		\begin{equation*}
			(d\psi)(x):=\langle \phi,\psi(x) \rangle,\quad
			x\in\mathbb{Z}, \quad 
			\psi\in \ell^2 (\mathbb{Z};\mathbb{C}^2),
		\end{equation*}
		then $d$ is a coisometry. 
		Forthermore, $d^\ast d$ has the form as follows and let $S$ be 
		\begin{equation*}
			d^\ast d =
			%\bigoplus_{x\in\mathbb{Z}} |\phi\rangle \langle \phi |\\
			%&=
			\bigoplus_{x\in\mathbb{Z}}
			\begin{pmatrix}
				|\phi_1 |^2 & \phi_1\overline{\phi_2} \\
				\overline{\phi_1} \phi_2  & |\phi_2 |^2
			\end{pmatrix},\quad % \\
			S:=
			\begin{pmatrix}
				p & qL \\
				\overline{q}L^\ast & -p
			\end{pmatrix}
			, 
		\end{equation*}
		where a pair $(p,q)\in \mathbb{R}\times\mathbb{C}$ satisfies $p^2 + |q|^2 = 1$ and $\phi_j$ denotes the $j$-th component of $\phi$ for $j=1,2$,
		with a direct decomposition 
		$\ell^2 (\mathbb{Z};\mathbb{C}^2)=\bigoplus_{x\in\mathbb{Z}}\mathbb{C}^2$.
		Then we have
		\begin{equation*}
			Sd^\ast d - d^\ast d S = 
			\begin{pmatrix}
				2i\Im(q\overline{\phi_1} \phi_2 L) & 2p\phi_1\overline{\phi_2} + q(|\phi_2|^2-|\phi_1|^2)L  \\
				-2p\overline{\phi_1}\phi_2 + \overline{q}(|\phi_1|^2-|\phi_2|^2)L^\ast &
				-2i\Im (q\overline{\phi_1} \phi_2 L)
			\end{pmatrix}
		\end{equation*}
		by the direct calculation, where $\Im A:=(A-A^\ast)/(-2i)$ is the imaginary part of a linear operator $A$. 
		From $\sigma(q\overline{\phi_1} \phi_2 L) = |q\phi_1 \phi_2|\mathbb{T}$, 
		$\Im (q\overline{\phi_1} \phi_2 L)$ vanishes only in the case $q\overline{\phi_1} \phi_2=0$. 
		Therefore we can show that $[S,d^\ast d]=\boldsymbol{0}$ if and only if $q=0$ and $\overline{\phi_1} \phi_2 = 0$. Also, the space-homogeneous quantum walk $U=SC$ is included in our model.
	\end{exam}
	
	We next give the example of a one-dimensional inhomogeneous non-unitary quantum walk.
	\begin{exam}\label{example3}
		Let $(\alpha ,\beta )\in (\mathbb{R},\mathbb{C})$. We set a operator $S\colon\ell^2\bigl(\mathbb{Z};\mathbb{C}^2\bigr)\to \ell^2\bigl(\mathbb{Z};\mathbb{C}^2\bigr)$ as 
		\begin{align*}
			S=\begin{pmatrix}
				0&L\\
				L^*&0
			\end{pmatrix},
		\end{align*} and two matrcies 
		\begin{align*}
			C_1=\begin{pmatrix}
				\alpha &\beta \\
				\beta^*&-\alpha
			\end{pmatrix},\quad 
			C_2=\begin{pmatrix}
				\lambda _+&0\\
				0&\lambda _-
			\end{pmatrix},
		\end{align*}where $\lambda _{\pm}=\sqrt{\alpha ^2+|\beta |^2}$. For $j=1,2$, let $\chi _j$ be the 
		normalized eigenvector of $C_j$ corresponding to $\lambda _+$. We define a coin operator $C$ on $\ell^2(\mathbb{Z};\mathbb{C}^2)$ as a multiplication operator by  
		\begin{align*}
			C(x)=\begin{cases}
				C_1,\quad &x:\text{odd,}\\
				C_2,&x:\text{even.}
			\end{cases}.
		\end{align*}
		Then $C$ is represented by $C=\lambda _+d^*d+\lambda _-(I-d^*d)$, where $d\colon\ell^2(\mathbb{Z};\mathbb{C}^2)\to \ell^2(\mathbb{Z})$ define as 
		\begin{align*}
			(d\Psi )(x)=\langle \chi (x),\Psi (x)\rangle ,\quad \Psi \in \ell^2\bigl(\mathbb{Z};\mathbb{C}^2\bigr)
		\end{align*} and 
		\begin{align*}
			\chi (x)=\begin{cases}
				\chi _1,\quad &x:\text{odd,}\\
				\chi _2,&x:\text{even.}
			\end{cases}.
		\end{align*}    
		Also, we have $[S,d^*d]\not =0$ by the direct calculation. Thus, the space-inhomogeneous quantum walk $U=SC$ is included in our model.
	\end{exam}
	\begin{comment}
	For a bounded operator $A$ on a Hilbert space $\mathcal{N}$, the symbols 
	\begin{align*}
	\sigma_\mathrm{r}(A)&:=\{z\in \mathbb{C} \mid A-z \ {\rm is}\ {\rm injective}\ {\rm and}\ \overline{ {\rm ran}(A-z)}\neq  \mathcal{N}\}\\
	\sigma_\mathrm{c}(A)&:=\{z\in\mathbb{C}\mid A-z \text{ is injective and } \overline{{\rm ran}(A-z)}=\mathcal{N}\neq{\rm ran}(A-z)\}\\
	\sigma_{\rm ap}(A)&:=\left\{z\in \mathbb{C} \ \middle| \ 
	\begin{array}{c} 
	{\rm there}\ {\rm exists}\ \{\psi_n\}_{n=1}^{\infty}\subset \mathcal{N}\ {\rm such}\ {\rm that}\\ \|\psi_n\|=1, n\in \mathbb{N} \ {\rm and} \ \lim_{n \rightarrow \infty} \|(A-z)\psi_n\| =0
	\end{array}
	\right\}
	\end{align*}
	are called respectively the residual spectrum, continuous spectrum and the approximate point spectrum of $A$. 
	\end{comment}
	\Acknowledgement{
		This work was supported by the Research Institute for Mathematical Sciences, an International Joint Usage/Research Center located in Kyoto
		University and the Grant-in-Aid of  Scientific Research (C) Japan Society for the Promotion of Science (Grant No.~19K03616).}
	
	%%%%%%%%%%%%%%%%%%%%%%%%%%%%%%%%%%%%%%%%%%%%%%%%%%%%%%%%%%%%%%%%%%%%%%%%%%%%
	
	%check memo:
	%how to write bibliography 
	
\end{document}